\documentclass[10pt]{llncs}
\usepackage{makeidx}  
\usepackage{amsfonts}
\usepackage{amssymb}
\usepackage{graphicx}
\usepackage{subfigure}
\usepackage[font=small]{caption}
\usepackage{wrapfig}
\usepackage{url}
\newcommand{\hlay}{\rightrightarrows}
\newcommand{\vlay}{\upuparrows}
\newcommand{\eqref}[1]{(\ref{#1})}

\makeatletter
\def\blfootnote{\xdef\@thefnmark{}\@footnotetext}
\makeatother

\begin{document}
\pagestyle{plain}
%
\title{Incremental Grid-like Layout\\ Using Soft and Hard Constraints}
\titlerunning{Incremental Grid-like Layout}  
%
\author{Steve Kieffer \and Tim Dwyer \and Kim Marriott \and Michael Wybrow}
\authorrunning{Steve Kieffer \emph{et al.}} 
%
\tocauthor{Steve Kieffer, Tim Dwyer, Kim Marriott, Michael Wybrow}
\institute{Caulfield School~of Information Technology,\\
Monash University, Caulfield, Victoria 3145, Australia,\\
National ICT Australia, Victoria Laboratory,\\
\email{\{Steve.Kieffer,Tim.Dwyer,Kim.Marriott,Michael.Wybrow\}@monash.edu}}

\maketitle              

\begin{abstract}
We explore various techniques to incorporate grid-like layout conventions into a force-directed, constraint-based graph layout framework.  In doing so we are able to provide high-quality layout---with predominantly axis-aligned edges---that is more flexible than previous grid-like layout methods and which can capture layout conventions in notations such as SBGN (Systems Biology Graphical Notation).  
Furthermore, the layout is easily able to respect user-defined constraints and adapt to interaction in online systems and diagram editors such as Dunnart.
\end{abstract}
\begin{figure}
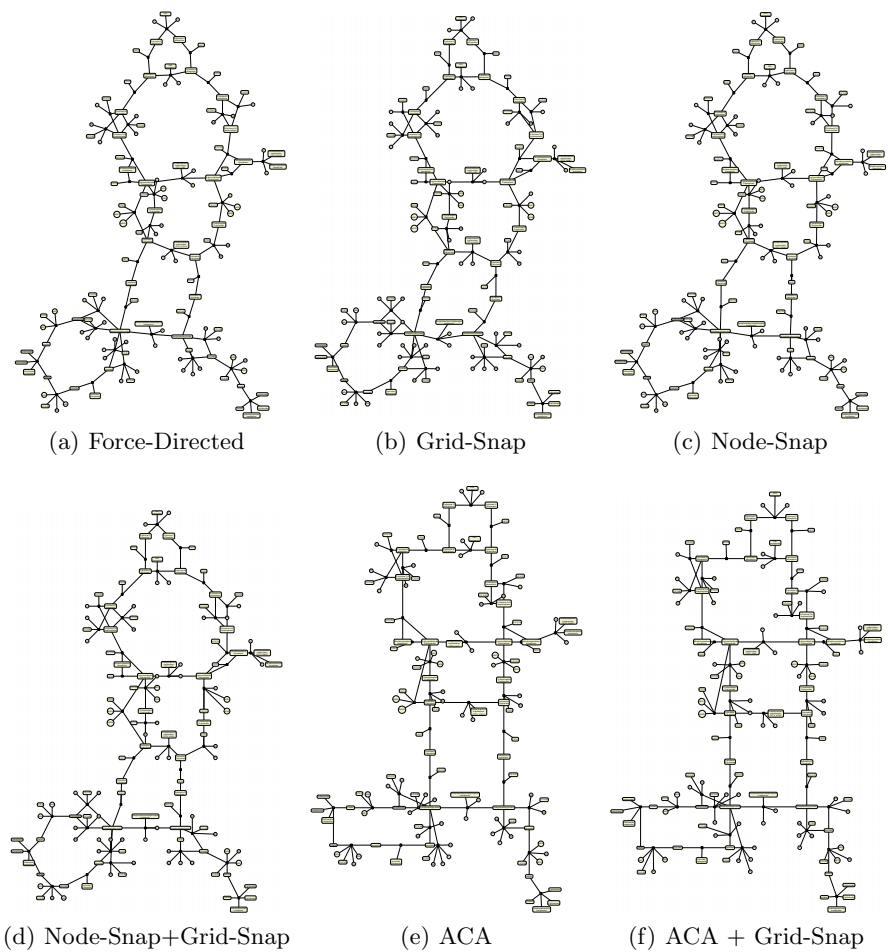

\centering
\subfigure[Force-Directed]{
\label{subfig:glyglu-FD}
\includegraphics[width=0.3\textwidth]{glyglu-FD2}
}
\subfigure[Grid-Snap]{
\label{subfig:glyglu-GS}
\includegraphics[width=0.3\textwidth]{glyglu-FD-OP-EN-GS2}
}
\subfigure[Node-Snap]{
\label{subfig:glyglu-NS}
\includegraphics[width=0.3\textwidth]{glyglu-FD-OP-NS2}
}
\subfigure[Node-Snap+Grid-Snap]{
\label{subfig:glyglu-NS-GS}
\includegraphics[width=0.3\textwidth]{glyglu-FD-OP-NS-GS2}
}
\subfigure[ACA]{
\label{subfig:glyglu-ACA}
\includegraphics[width=0.3\textwidth]{glyglu-FD-ACA2}
}
\subfigure[ACA + Grid-Snap]{
\label{subfig:glyglu-ACA-GS}
\includegraphics[width=0.3\textwidth]{glyglu-FD-ACA-OP-EN-GS2}
}
\caption{
Different combinations of our automatic layout techniques for grid-like layout compared 
with standard force-directed layout.
The layout is for an SBGN (Systems Biology Graphical Notation) diagram
of the Glycolysis-Gluconeogenesis pathway obtained from MetaCrop
\cite{metacrop}.
In SBGN  diagrams, \emph{process nodes} represent individual chemical
reactions which typically form links in long metabolic pathways, and
are often connected to several degree-1 nodes representing
``currency molecules'' like ATP and ADP, while precisely two of their
neighbours are degree-2 nodes representing principal metabolites.
It is conventional that the edges connecting main chemicals and process
nodes be axis-aligned in long chains, but not the leaf edges.
We achieve this by tailoring the cost functions
discussed in \S\ref{sec:constraint}.
}
\label{fig:Glycolysis}
\end{figure}

\section{Introduction}
Force-directed layout remains the most popular approach to automatic layout of undirected graphs. By and large these metheds untangle the graph to show underlying structure and symmetries with a layout style that is organic in appearance~\cite{battista1998graph}. Constrained graph layout methods extend force-directed layout to take into account user-specified constraints on node positions such as alignment, hierarchical containment and non-overlap~\cite{dwyer2006ipsep}. These methods have proven a good basis for semi-automated graph layout in tools such as Dunnart~\cite{dwyer2009dunnart} that allow the user to interactively guide the layout by moving nodes or adding constraints.
\blfootnote{A version of this paper has been accepted for publication in Graph Drawing 2013.
The final publication will be available at \url{link.springer.com}.}

However, when undirected graphs (and other kinds of diagrams) are drawn by hand it is common for a more grid-like layout style to be used. Grid-based layout is widely used by graphic designers and it is common in hand-drawn biological networks and metro-map layouts.  Previous research has shown that grid-based layouts are more memorable than unaligned placements~\cite{marriott2012memorability}.
Virtually all diagram creation tools provide some kind of snap-to-grid feature.

In this paper we investigate how to modify constrained force-directed graph layout
methods \cite{dwyer2006ipsep} to create more orthogonal and grid-like layouts with a particular focus
on interactive applications such as Dunnart.
In Figure \ref{fig:Glycolysis}
we show undirected graphs
arranged with our various layout approaches compared with traditional force-directed layout. 

Before proceeding, it is worth defining what we mean by a \emph{grid-like layout}.
It is commonly used to mean some combination of the following properties: 
\begin{enumerate}
\item nodes are positioned at points on a fairly coarse grid;
\item edges are simple horizontal or vertical lines or in some cases 45$^\circ$ diagonals;
\item nodes of the same kind are horizontally or vertically aligned;
\item edges are orthogonal, i.e.,\ any bends are $90^\circ$.
\end{enumerate}
and thus is different from the notion of a \emph{grid layout}, which is simply property (1).
In this paper we are primarily interested in producing layouts with properties (1) and (2), though our methods could also achieve (3). We do not consider edges with orthogonal bends, though this could be an extension or achieved through a routing post-process (a simple example of this is provided in the Appendix).

The standard approach to extending force-directed methods to handle new aesthetic criteria is to add extra ``forces'' which push nodes in order to satisfy particular aesthetics.  
One of the most commonly used functions is \emph{stress}~\cite{gansner2005graph}. Our first contribution ($\S$\ref{sec:force}) is to develop penalty terms that can be added to the stress function to reward placement on points in a grid (Property 1) and to
reward horizontal or vertical node alignment and/or horizontal or vertical edges (Property 2 or 3).
We call these the \emph{Grid-Snap} and \emph{Node-Snap} methods respectively.


However, 
additional terms can make the goal function rich in local minima that impede convergence to a more aesthetically pleasing global minimum.  Also, such ``soft'' constraints cannot guarantee satisfaction and so layouts in which nodes are \emph{nearly-but-not-quite aligned} can occur. For this reason we investigate a second approach based on constrained graph layout in which \emph{hard} alignment constraints are automatically added to the layout so as to ensure horizontal or vertical node alignment and thus horizontal or vertical edges (Property 2 or 3).
This \emph{adaptive constrained alignment (ACA)} method ($\S$\ref{sec:constraint}) is the most innovative contribution of our paper.

In $\S$\ref{sec:eval} we provide an empirical investigation of the speed of these approaches and  the quality of layout with respect to various features encoding what we feel are the aesthetic criteria important in grid-like network layout. 

While the above approaches can be used in once-off network layout, our original motivation was for interactive-layout applications. In $\S$\ref{sec:interactive} we discuss an interaction model based on the above for the use of grid-like layout in interactive semi-automatic layout tools such as Dunnart.


\vspace{2mm}
\noindent\textbf{Related Work:} Our research is related to proposals for automatic grid-like layout of biological networks~\cite{barsky2007cerebral,li2005grid,kojima2007efficient}. These arrange biological networks with grid coordinates for nodes in addition to various layout constraints.  In particular Barsky \emph{et al.}~\cite{barsky2007cerebral} consider alignment constraints between biologically similar nodes and Kojima~\emph{et al.}~\cite{kojima2007efficient} perform layout subject to rectangular containers around functionally significant groups of nodes (e.g.,\ metabolites inside the nucleus of a cell).  In general they use fairly straight-forward simulated annealing or simple incremental local-search strategies.  Such methods work to a degree but are slow and may never reach a particularly aesthetically appealing minimum.

Another application where grid-like layout is an important aesthetic is automatic metro-map layout.
Stott \emph{et al.}~\cite{stott2011automatic} use a simple local-search (``hill-climbing'') technique to obtain layout on grid points subject to a number of constraints, such as octilinear edge orientation.  Wang and Chi \cite{wang2011focus} seek similar layout aesthetics but using continuous non-linear optimization subject to octilinearity and planarity constraints.  This work, like ours, is based
on a quasi-Newton optimization method, but it is very specific to metro-map layout and it is not at all clear how
these techniques could be adapted to general-purpose interactive diagramming applications.

Another family of algorithms that compute grid-like layout are so-called \emph{orthogonal} graph drawing methods.  There have been some efforts to make these incremental, for example Brandes \emph{et al.}~\cite{brandes2002sketch} can produce an orthogonal drawing of a graph that respects the topology for a given set of initial node positions.  Being based on the ``Kandinski'' orthogonal layout pipeline, extending such a method with user-defined constraints such as alignment or hierarchical containment would require non-trivial engineering of each stage in the pipeline.  There is also a body of theoretical work considering the computability and geometric properties of layout with grid-constraints for various classes of graphs, e.g.~\cite{chrobak1995linear}.  Though interesting in its own right, such work is usually not intended for practical application, which is the primary concern of this paper.

There are several examples of the application of soft-constraints to layout.
Sugiyama and Misue~\cite{sugiyama1995graph} augment the standard force-model with ``magnetic'' edge-alignment forces.  Ryall \emph{et al.}~\cite{ryall1997interactive} explored the use of various force-based constraints in the context of an interactive diagramming editor.  It is the limitations of such soft constraints (discussed below) which
prompt the development of the techniques described in
\S\ref{sec:constraint}.

\section{Aesthetic Criteria}\label{sec:aesthetic}

Throughout this paper we assume that we have a graph $G = (V,E,w,h)$ consisting of a set of nodes $V$, a set of edges $E \subseteq V \times V$ and
$w_v, h_v$ are the width and height of node $v \in V$.
We wish to find a straight-line  2D drawing for $G$. This  is specified by a pair $(x,y)$ where
$(x_v, y_v)$ is the centre point of each $v \in V$.

We quantify grid-like layout quality through the following metrics.  
In subsequent sections we use these to develop soft and hard constraints that directly or indirectly aim to optimise them.  
We also use these metrics in our evaluation $\S$\ref{sec:eval}.

\vspace{1.5mm}
\noindent\textbf{Embedding quality} We measure this using the \emph{P-stress} function~\cite{dwyer2009topology}, a variant of \emph{stress}~\cite{gansner2005graph} that does not penalise unconnected nodes being more than their desired distance apart. It measures the separation between each pair of nodes $u, v \in V$ in the drawing and their \emph{ideal distance} $d_{u v}$ proportional to the graph theoretic path between them:
{\small
$$
  \sum_{u < v \in V} w_{u v}
  \left( \left( d_{u v} - d(u,v) \right)^{+} \right)^{2} +
  \sum_{(u,v) \in E} wp \left( \left( d(u,v) - d_L   \right)^{+} \right)^{2}
$$
where $d(u,v)$ is the Euclidean distance between $u$ and $v$, $(z)^+=z$ if $z \ge 0$ otherwise $0$, $d_L$ is an ideal edge length, $wp = \frac{1}{d_L}$, and $w_{u v} = \frac{1}{d_{u v}^2}$.
}

\noindent\textbf{Edge crossings} The number of edge crossings in the drawing.

\noindent\textbf{Edge/node overlap} The number of edges intersecting a node box.  With straight-line edges this also penalises coincident edges.\footnote{Node/node overlaps are also undesirable. We avoid them completely by using hard non-overlap constraints~\cite{dwyer2006fast} in all our tests and examples.}

\noindent\textbf{Angular resolution}  Edges incident on the same node have a uniform  angular separation.  Stott \emph{et al.}~\cite{stott2011automatic} give a useful formulation:
$$
\sum_{v \in V} \sum_{\{e_1,e_2\}\in E } \left|2\pi/\mathit{degree}(v)-\theta(e_1,e_2)\right|
$$

\noindent\textbf{Edge obliqueness} We prefer horizontal or vertical edges and then---with weaker preference---edges at a 45$^\circ$ orientation.  Our precise metric is 
$
  M \left|\tan^{-1} \frac{y_u-y_v}{x_u-x_v}\right|
$
where $M(\theta)$ is an ``M-shaped function''
over $[0,\pi/2]$ 
that highly penalizes edges which are almost but not quite axis-aligned and gives a lower penalty for edges midway between horizontal and vertical.\footnote{Note that $[0,\pi/2]$ is the range of $|tan^{-1}|$.
The ``M'' function is zero at $0$ and $\pi/2$, a small value $p \geq 0$ at $\pi/4$, a large value $P > 0$
at $\delta$ and $\pi/2-\delta$ for some small $\delta > 0$, and linear in-between.}
Other functions like those of \cite{stott2011automatic,kojima2007efficient} could be used instead.

\noindent\textbf{Grid placement} Average of distances of nodes from their closest grid point.

\section{Soft-Constraint Approaches}\label{sec:force}

In this section we describe two new terms that can be combined with the \emph{P-stress} function to
achieve more grid-like layout: \emph{NS-stress} for ``node-snap
stress'' and \emph{GS-stress} for ``grid-snap stress.''  An additional term \emph{EN-sep} gives good separation between nodes and edges.
Layout is then achieved by minimizing 
$$\mbox{P-stress} + k_{ns}\cdot\mbox{NS-stress} + k_{gs}\cdot\mbox{GS-stress}+k_{en}\cdot\mbox{EN-sep}$$ 
where
$k_{ns,gs,en}$ control the ``strength'' of the various components.  
These extra terms, as defined below, tend to make nodes lie on top of
one another. It is essential to avoid this by solving subject to node-overlap
prevention constraints, as described in \cite{dwyer2006fast}.  To obtain an initial ``untangled'' layout we run with $k_{ns}=k_{gs}=k_{en}=0$ and without non-overlap constraints (Fig.~\ref{subfig:glyglu-FD}),
and then run again with the extra terms and constraints to perform ``grid beautification''.




Minimization of the \emph{NS-stress} term favours horizontal or vertical alignment of pairs of connected nodes (Figs.~\ref{subfig:glyglu-NS} and~\ref{fig:CalvinCycle}).
Specifically, taking $\sigma$ as the distance at which nodes should snap into
alignment with one another, we define:

{\small
\[
  \mbox{NS-stress} = 
     \sum_{(u,v) \in E} q_\sigma(x_u-x_v) + q_\sigma(y_u-y_v)
     ~~~\mbox{where}~
  q_\sigma(z) = \left\lbrace\begin{array}{cl}
    z^2/\sigma^2 & \, |z| \leq \sigma \\
    0 & \, \mathrm{otherwise.}
  \end{array}\right.
\]
}


We originally tried several
other penalty functions which turned out not to have good convergence.
In particular any smooth function with local maxima at $\pm\sigma$
must be concave-down somewhere over the interval $[-\sigma,\sigma]$,
and while differentiability may seem intuitively desirable for quadratic
optimization it is in fact trumped by downward concavity, which plays
havoc with standard step-size calculations on which our
gradient-projection algorithm is based.
Thus, obvious choices like
an inverted quartic $(1+(z^2-\sigma^2)^2)^{-1}$
or a sum of inverted quadratics
$(1+(z+\sigma)^2)^{-1} + (1+(z-\sigma)^2)^{-1}$
proved unsuitable in place
of $q_\sigma(z)$. We review the step size, gradient, and Hessian
formulae for our snap-stress functions in the Appendix.

We designed our \emph{GS-stress} function likewise
to make the lines
of a virtual grid exert a similar attractive force on nodes once
within the snap distance $\sigma$:
\[
  \mbox{GS-stress} = 
    \sum_{u \in V} q_\sigma(x_u-a_u) + q_\sigma(y_u-b_u)
\]
where $(a_u,b_u)$ is the closest grid point to $(x_u,y_u)$ (with ties
broken by favouring the point closer to the origin), see Fig.~\ref{subfig:glyglu-GS}.
The grid is defined to be the set of all points $(n\tau, m\tau)$, where $n$ and $m$
are integers, and $\tau$ is the ``grid size''.  With \emph{GS-stress} active it is important to set some other parameters
proportional to $\tau$.  First, we take $\sigma = \tau/2$.
Next, we modify the non-overlap constraints to allow no more than one node centre
to be in the vicinity of any one grid point by increasing the minimum separation distance allowed between adjacent nodes to $\tau$.
Finally, we found that setting the ideal
edge length equal to $\tau$ for initial force-directed
layout, before activating \emph{GS-stress}, helped to put
nodes in positions compatible with the grid.

Our third term \emph{EN-sep} is also a quadratic function based on $q_\sigma(z)$
that separates nodes and nearby axis-aligned edges to avoid node/edge overlaps and coincident edges:
\[
  \mbox{EN-sep}=\sum_{e \in E_V \cup E_H} \sum_{u \in V}
     q_{\sigma}\left( (\sigma-d(u,e))^+ \right),
\]
where 
$E_V$ and $E_H$ are the sets of vertically and horizontally
aligned edges, respectively,
and the distance
$d(u,e)$
between a
node $u$ and an edge $e$ is defined as the length of the normal from
$u$ to $e$ if that exists, or $+\infty$ if it does not.
Here again we took $\sigma = \tau/2$.

\smallskip
\noindent
\framebox{
\parbox{0.95\textwidth}{
\small
In our experiments we refer to various combinations of these terms and constraints:

\noindent {\bf Node-Snap:} \emph{NS-stress}, \emph{EN-sep}, non-overlap constraints, $k_{gs}=0$

\noindent {\bf Grid-Snap:} \emph{GS-stress}, \emph{EN-sep}, ideal edge lengths equal to grid size, non-overlap,
constraints with separations tailored to grid size, $k_{ns}=0$.

\noindent {\bf Node-Snap+Grid-Snap:} achieves extra alignment by adding \emph{NS-stress} to the above {\bf Grid-Snap} recipe (i.e. $k_{ns}\ne 0$)
}
}

\section{Adaptive Constrained Alignment}\label{sec:constraint}


%
%
%

Another way to customize constrained force-directed layout is by
adding \emph{hard} constraints, and in this section we describe how to make
force-directed layouts more grid-like simply by adding
alignment and separation constraints (Fig.~\ref{subfig:glyglu-ACA}). 

The algorithm, which we call \emph{Adaptive Constrained Alignment or ACA},
 is 
a greedy algorithm which repeatedly chooses an edge in $G$ and aligns
it horizontally or vertically
(see $\mathit{adapt\_const\_align}$ procedure of Figure~\ref{fig:ACAsubprocs}).
It \emph{adapts} to user specified constraints by not adding alignments that violate
these.  
The algorithm halts when the heuristic can no longer
apply alignments without creating edge overlaps.
Since each edge is aligned at most once, there are at
most $\left|E\right|$ iterations.

We tried the algorithm with three different heuristics for choosing
potential alignments, which we discuss
below.




Node overlaps and edge/node overlaps can be prevented with hard
non-overlap constraints and the \emph{EN-sep} soft constraint discussed in Section~\ref{sec:force},
applied either before
or after the ACA process. However, 
coincident edges can be accidentally created and then enforced as we apply
alignments if we do not take care to maintain the orthogonal ordering of nodes.
If for example two edges $(u, v)$ and $(v, w)$ sharing a common
endpoint $v$ are both horizontally aligned, then we must maintain
either the ordering $x_u < x_v < x_w$ or the opposite
ordering $x_w < x_v < x_u$.  

Therefore we define the notion of a \emph{separated alignment},
written $\mathsf{SA}(u,v,D)$ where $u,v \in V$
and $D \in \{\mathbb{N}, \mathbb{S}, \mathbb{W}, \mathbb{E} \}$ is a
compass direction. Applying a separated alignment means applying two
constraints to the force-directed layout---one alignment and one
separation---as follows:
\[
\begin{array}{rccccrcl}
\mathsf{SA}(u,v,\mathbb{N}) \equiv & x_u = x_v & \mathrm{and} &
                               y_v + \beta(u,v) \leq y_u, &
    \quad &
    \mathsf{SA}(u,v,\mathbb{S}) & \equiv & \mathsf{SA}(v,u,\mathbb{N}), \\

\mathsf{SA}(u,v,\mathbb{W}) \equiv & y_u = y_v & \mathrm{and} &
                               x_v + \alpha(u,v) \leq x_u, &
    \quad &
    \mathsf{SA}(u,v,\mathbb{E}) & \equiv & \mathsf{SA}(v,u,\mathbb{W}),
\end{array}
\]
where $\alpha(u,v) = (w_u+w_v)/2$ and $\beta(u,v) = (h_u+h_v)/2$.
\label{page:alphaBetaDef}
(Thus for example $\mathsf{SA}(u,v,\mathbb{N})$ can be read as, ``the
ray from $u$ through $v$ points north,'' where we think of $v$ as lying
north of $u$ when its $y$-coordinate is smaller.)


\begin{figure}
\noindent
\parbox[t]{0.5\textwidth}{
{\scriptsize
{\bf proc} $\mathit{adapt\_const\_align}(G,C,H)$

\quad $(x,y) \leftarrow \mathit{cfdl}(G,C)$

\quad $\mathit{SA} \leftarrow H(G,C,x,y)$

\quad $\mathbf{while} \: \mathit{SA} \: != \: \mathit{NULL}$

\qquad  $C.\mathit{append}(\mathit{SA})$

\qquad  $(x,y) \leftarrow \mathit{cfdl}(G,C)$



\qquad  $\mathit{SA} \leftarrow H(G,C,x,y)$

\quad $\mathbf{return} \: (x,y,C)$
}
}
\parbox[t]{0.5\textwidth}{
{\scriptsize
{\bf proc} $\mathit{chooseSA}(G,C,x,y,K)$

\quad $S \leftarrow NULL$

\quad $\mathit{cost} \leftarrow \infty$

\quad {\bf for each} $(u,v) \in E$ {\bf and} dir. $D$

\qquad  {\bf if not} $\mathit{creates\_coincidence}(C,x,y,u,v,D)$

\qquad \quad  {\bf if} $K(u,v,D) < \mathit{cost}$

\qquad \qquad   $S \leftarrow \mathit{SA}(u,v,D)$

\qquad \qquad   $\mathit{cost} \leftarrow K(u,v,D)$

\quad {\bf return} $S$
}
}
\caption{Adaptive constrained alignment algorithm.
$G$ is the given graph, $C$ the set of user-defined
constraints, $H$ the alignment choice heuristic, and \emph{cfdl} the
constrained force-directed layout procedure.
}
\label{fig:ACAsubprocs}
\end{figure}

%

\subsubsection{Alignment Choice Heuristics.}
We describe two kinds of alignment choice heuristics: \emph{generic},
which can be applied to any graph, and \emph{convention-based}, which
are intended for use with layouts that must conform to special
conventions, for example SBGN diagrams~\cite{lenovere2009sbgn}.
All of our heuristics are designed according to two principles:
\begin{enumerate}
\item Try to retain the overall shape of the initial force-directed layout.

\item Do not obscure the graph structure by creating undesirable overlaps.
\end{enumerate}
and differ only in the choice of a \emph{cost
function} $K$ which is plugged into the procedure {\tt chooseSA}
in Figure \ref{fig:ACAsubprocs}. This relies on procedure
{\tt creates\_coincidence} which implements the edge coincidence test
described by Theorem 1.
Among separated alignments which would not lead to an edge
coincidence, {\tt chooseSA} selects one of lowest cost.
Cost functions may return a special value of $\infty$ to mark an
alignment as never to be chosen.

The {\tt creates\_coincidence} procedure works by maintaining a
$|V|$-by-$|V|$ array of flags which indicate for
each pair of nodes $u, v$ whether they are aligned in either dimension
and whether there is an edge between them.
The cost of initializing the array is
$\mathcal{O}(|V|^2 + |E| + \left|C\right|)$,
but this is done only once in ACA.
Each time a new alignment constraint is added
the flags are updated in $\mathcal{O}(|V|)$ time, due to transitivity of
the alignment relation.
Checking whether a proposed separated alignment would create an
edge coincidence also takes $\mathcal{O}(|V|)$ time, and works
according to Theorem~1. (Proof is provided in the Appendix.)
Note that the validity of Theorem~1 relies on the fact that
we apply separated alignments $\mathsf{SA}(u,v,D)$ only when
$(u,v)$ is an edge in the graph.

\medskip
\noindent {\bf Theorem 1.}
\emph{Let $G$ be a graph with separated alignments.
Let $u, v$ be nodes 
in $G$ which are not yet constrained to one another.
Then the separated alignment
$\mathsf{SA}(u,v,\mathbb{E})$ creates an edge coincidence in $G$ if and only
if there is a node $w$
which is horizontally aligned with either $u$ or $v$ and
satisfies either of the following two conditions:
(i)  $(u,w) \in E$ while $x_u < x_w$ or $x_v < x_w$; or
(ii) $(w,v) \in E$ while $x_w < x_v$ or $x_w < x_u$.
%
%
The case of vertical alignments is similar.
}

\medskip
We tried various cost functions, which addressed the aesthetic criteria of
Section~\ref{sec:aesthetic} in different ways.
We began with a \emph{basic cost}, which was either an estimate
$K_{dS}(u,v,D)$ of the change in the
stress function after applying the proposed alignment
$\mathsf{SA}(u,v,D)$, or else the negation of the obliqueness of the edge,
$K_{ob}(u,v,D) = -obliqueness((u,v))$, as measured by the function of
Section~\ref{sec:aesthetic}.
In this way we could choose to address
the aesthetic criteria of \emph{embedding quality} or
\emph{edge obliqueness},
and we found that the results were similar. Both rules favour placing
the first alignments on edges which are almost axis-aligned, and this
satisfies our first principle of being guided as much as possible by
the shape of the initial force-directed layout.
See for example Figure~\ref{fig:Glycolysis}.

On top of this basic cost
we considered \emph{angular resolution} of degree-2 nodes
by adding a large but finite cost that would postpone certain
alignments until after others had been attempted;
namely,
we added a fixed cost
of 1000 (ten times larger than average values of $K_{dS}$ and $K_{ob}$) for any alignment that would make a degree-2 node into a
``bend point,'' i.e.,~would make one of its edges horizontally aligned
while the other was vertically aligned.
This allows long chains of degree-2 nodes to form
straight lines, and cycles of degree-2 nodes to form
perfect rectangles.
For SBGN diagrams we
used a modification of this rule based on \emph{non-leaf degree}, or number
of neighbouring nodes which are not leaves (Figs.~\ref{subfig:glyglu-ACA} and~\ref{subfig:glyglu-ACA-GS}).

\subsubsection{Respecting User-Defined Constraints.}

Layout constraints can easily wind up in conflict with one another if
not chosen carefully.
In Dunnart such conflicts are detected during the 
projection operation described in~\cite{dwyer2006ipsep}, an \emph{active set}
method which iteratively determines the most violated constraint $c$ and satisfies it
by minimal disturbance of the node positions. When it is impossible to
satisfy $c$ without violating one of the constraints that is already
in the active set, $c$ is simply marked \emph{unsatisfiable}, 
and the operation carries on without it.

For ACA it is important that user-defined constraints are never marked
unsatisfiable in deference to an alignment imposed by the process;
therefore we term the former \emph{definite} constraints and the
latter \emph{tentative} constraints.  We employ a modified projection
operation which always chooses to mark one or more tentative constraint as
unsatisfiable if they are involved in a conflict.

For conflicts involving more than one tentative constraint, we use
Lagrange multipliers to choose which one to reject. These are computed as a part of the projection process.
Since alignment constraints are
equalities (not inequalities) the sign of their Lagrange multiplier
does not matter, and a constraint whose Lagrange multiplier is
maximal in absolute value is one whose rejection should permit the
greatest decrease in the stress function. Therefore we choose this one.


ACA does not snap nodes to grid-points: if desired this can be achieved
once ACA has added the alignment constraints by
activating Grid-Snap.

%
%

\section{Interaction}\label{sec:interactive}
\begin{figure}[t]
\centering
\includegraphics[trim=0em 0em 0em 0em,width=0.75\textwidth]{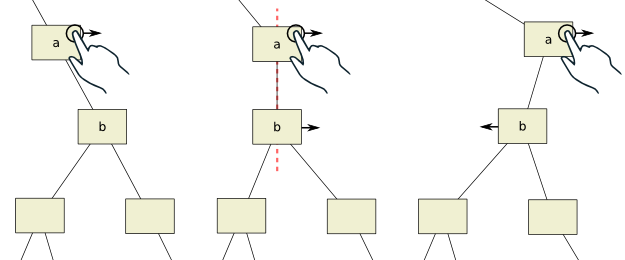}
\caption{
Interacting with Node-Snap.
The user is dragging node $a$ steadily to the right.
When the horizontal distance between $a$ and $b$ is less than the average width of these two nodes,
the \emph{NS-stress} function causes $b$ to align with $a$.
As the user continues dragging, the now aligned node $b$ will follow
until either a quick jerk of node $a$ breaks the alignment, or else edges attached to $b$
pull it back to the left, overcoming its attraction to $a$.
To the user, the impression is that the
alignment persisted until it was ``torn'' by the underlying forces in the system.
\label{fig:adaptivedragging}}
\end{figure}
One benefit of the approaches described above is that they are immediately applicable for use in interactive tools where the underlying graph, the prerequisite constraint system, or ideal
positions for nodes can all change dynamically.
We implemented Node-Snap, Grid-Snap and Adaptive Constrained Alignment
for interactive use in the Dunnart diagram editor.\footnote{
A video demonstrating interactive use of the approaches
described in this paper is available at
\url{http://www.dunnart.org}.
}
In Dunnart, automatic layout runs continuously in a background worker thread, allowing the layout to adapt immediately to user-specified changes to positions or constraints.

For example, Figure \ref{fig:adaptivedragging} illustrates user interaction with Node-Snap.
As the user drags a node around the canvas, it may snap into alignment with an adjacent
node. Slowly dragging a node aligned with other nodes will move them together
and keep them in alignment, while quickly dragging a node will instead cause it to be
torn from any alignments.

When we tried Node-Snap interactively in Dunnart we found that
nodes tended to stick together in clumps if the $\sigma$ parameter of \emph{NS-stress} was
larger than their average size in either dimension. We solved this
problem by replacing the snap-stress term by
\[
   \sum_{(u,v) \in E} q_{\alpha(u,v)}(x_u-x_v) +
                       q_{\beta(u,v)}(y_u-y_v)
\]
where $\alpha, \beta$ are as on page \pageref{page:alphaBetaDef}.


In Dunnart, a dragged object is always pinned to the mouse cursor.
In the case of Grid-Snap, the dragged node is unpinned and will immediately snap to a grid point on mouse-up.
Other nodes, however, will snap-to or tear-away from grid points in response to changing dynamics in the layout system.
During dragging we also turn off non-overlap constraints and reapply them on mouse-up.
This prevents nodes being unexpectedly pushed out of place as a result of the expanded non-overlap region ($\S$\ref{sec:force}).
Additionally, since \emph{GS-stress} holds nodes in place, we allow the user to quickly drag a
node to temporarily overcome the grid forces and allow the layout to untangle with
standard force-directed layout.  Once it converges we automatically reapply \emph{GS-stress}.

\section{Evaluation}\label{sec:eval} 

To evaluate the various techniques we applied each to 252 graphs from the ``AT\&T Graphs'' corpus (\url{ftp://ftp.research.att.com/dist/drawdag/ug.gz}) with between 10 and 244 nodes.  We excluded graphs with fewer than 10 nodes
 and two outlier graphs: one with 1103 nodes and one with 0 edges.
We recorded running times of each stage in the automated batch process and
the various aesthetic metrics described in $\S$\ref{sec:aesthetic},
using
a MacBook Pro with a 2.3GHz Intel Core I7 CPU.
Details of collected data etc.\ are given in the Appendix.

We found that ACA was the slowest approach, often taking up to 10 times as long as the other methods, on average around 5 seconds for graphs with around 100 nodes, while the other approaches took around a second.  ACA was also sensitive to the density of edges.  Of the soft constraint approaches, Grid-Snap (being very local) added very little time over the unconstrained force-directed approach.  

\begin{figure}[t]
\centering
\begin{minipage}[t]{0.45\textwidth}
\centering
\includegraphics[width=\linewidth]{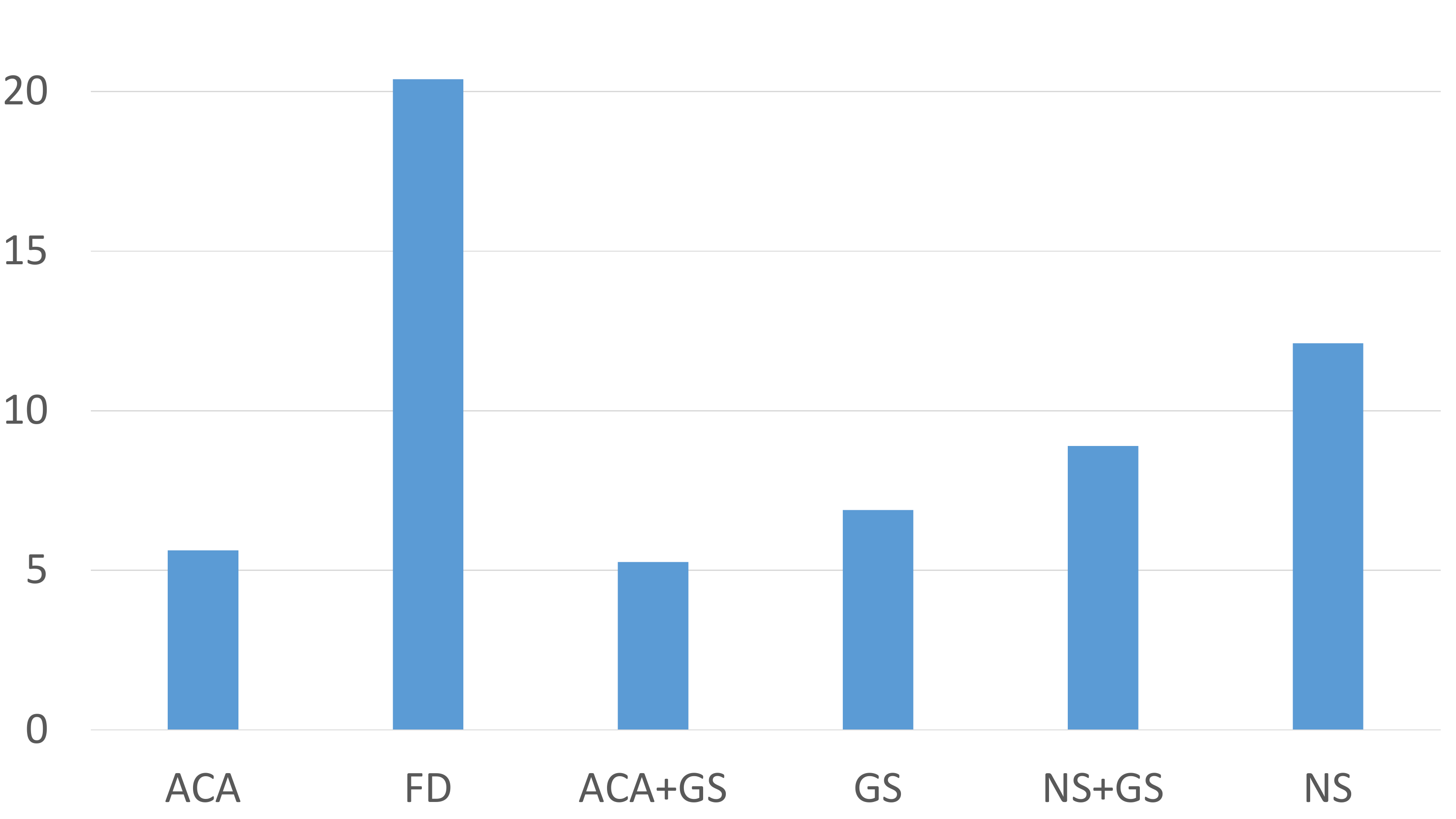}
\captionof{figure}{Edge obliqueness (see $\S$\ref{sec:aesthetic}) results.  The hard-constraint approach ACA 
is better than either of the soft constraint approaches Grid-Snap (GS) and Node-Snap (NS).  The combination of ACA and GS gives the best result.
\label{fig:obliqueness}}
\end{minipage}~~~
\begin{minipage}[t]{0.48\textwidth}
\centering
\includegraphics[width=0.48\textwidth]{Calvin-FD-OP-NS}
 \includegraphics[width=0.45\textwidth]{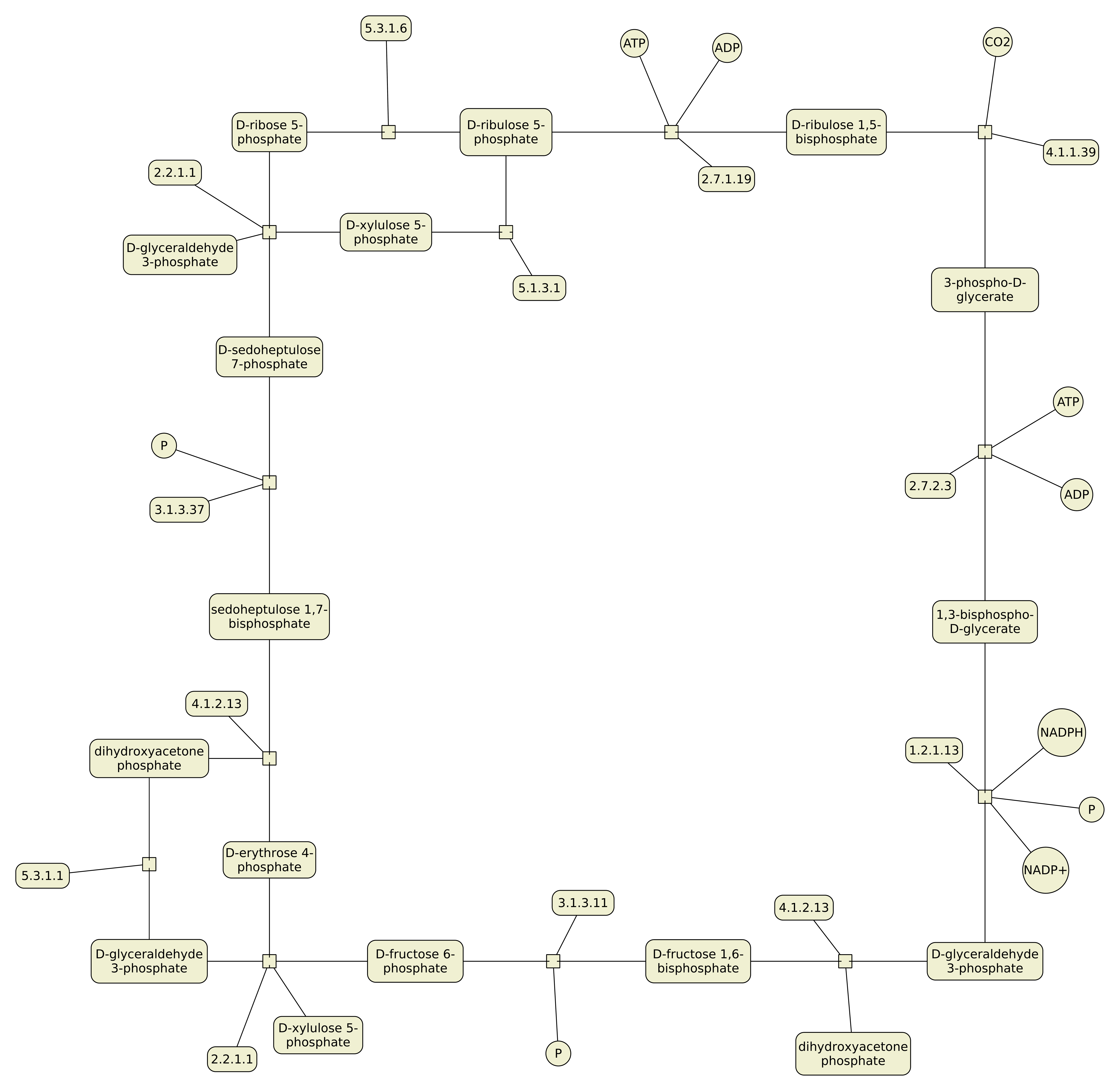}
\captionof{figure}{Layout of a SBGN diagram of Calvin Cycle pathway shows how ACA (right) gives a more pleasing rectangular layout than Node-Snap (left).}
\label{fig:CalvinCycle}
\end{minipage}
\end{figure}

The Edge Obliqueness (see $\S$\ref{sec:aesthetic}) results are shown in Fig. \ref{fig:obliqueness} as this is arguably the metric that is most indicative of grid-like layout.  Another desirable property of grid-like layout, as noted in $\S$\ref{sec:constraint} is that longer paths in the graph also be aligned.  ACA does a good job of aligning such paths, as is visible in Fig.\ \ref{fig:Glycolysis} and~\ref{fig:CalvinCycle}.

\section{Conclusion}
We have explored how to incorporate grid-like layout conventions into a force-directed,
constraint-based graph layout framework. We give two \emph{soft} approaches
(Node-Snap, Grid-Snap)
based on adding terms to the goal function, and an adaptive constraint based approach (ACA)
in which \emph{hard} alignment constraints are added greedily.
We find the ACA approach is slower but gives more grid-like layout and so
is the method of choice for once-off layout, at least for medium sized graphs.

We have also discussed how the approaches can be integrated into interactive diagramming tools like Dunnart. 
For interactive use both ACA and Grid-Snap provide good initial layouts, while Node-Snap helps the user create further alignments by hand.

Future work is to improve the speed of ACA by adding more than one alignment constraint at a time
and also to use Lagrange multipliers to improve the adaptivity of ACA.
One idea is to automatically reject any alignment whose Lagrange multiplier exceeds a predetermined threshold
on each iteration of ACA.  With this extension, running ACA continuously during interaction would allow us to achieve the behaviour
illustrated in Fig.~\ref{fig:adaptivedragging} through hard rather than soft constraints.
Another issue with all the techniques described is the many fiddly parameters, weights and thresholds.
We intend to further investigate principled ways to automatically set these.



%
%
\bibliographystyle{splncs03}
\bibliography{GridLayout}

\appendix
\section{Appendix}

\subsection{Gradient-projection}

The descent step of our gradient-projection algorithm computes
the descent direction and step size in terms of the gradient
and Hessian (matrix of mixed second partial derivatives) of our stress
function $S$. Namely, if $g = \nabla S$ and $H = \nabla^2 S$ then
the descent direction is $-g$ and the step size is
\[
    \frac{g^T g}{g^T H g}.
\]
(See for example \cite{NocedalWright} p.~47.)

The terms in $g$ and $H$ corresponding to \emph{P-stress} are
given in \cite{dwyer2009topology}.
Here we give the terms corresponding to the following three
functions,
\begin{eqnarray*}
N & = & \sum_{(u,v) \in E} q_{\sigma}(x_u-x_v) +
                       q_{\sigma}(y_u-y_v) \\
G & = & \sum_{u \in V} q_\sigma(x_u-a_u) + q_\sigma(y_u-b_u) \\
E & = & \sum_{e \in E_V \cup E_H} \sum_{u \in V}
     q_{\sigma}\left( (\sigma-d(u,e))^+ \right),
\end{eqnarray*}
which are the node-snap, grid-snap, and edge-node repulsion terms
from Section~3, respectively.
For $\sigma > 0$ we define
\[
    \gamma_\sigma(z) =
    \left\lbrace\begin{array}{cl}
    2z/\sigma^2 & \quad |z| \leq \sigma \\
    0 & \quad \mathrm{otherwise}
  \end{array}\right.
\]
and
\[
    \eta_\sigma(z) =
    \left\lbrace\begin{array}{cl}
    2/\sigma^2 & \quad |z| \leq \sigma \\
    0 & \quad \mathrm{otherwise.}
  \end{array}\right.
\]

For node-snap forces we have
\[
    \frac{\partial N}{\partial x_u} =
    \sum_{(u,v)\in E} \gamma_{\sigma}(x_u-x_v)
\]
and
\[ 
    \frac{\partial^2 N}{\partial x_v \partial x_u} =
        \left\lbrace\begin{array}{cl}
            -\eta_{\sigma}(x_u-x_v) & \quad\mathrm{if}\: (u,v) \in E \\
            0 & \quad\mathrm{otherwise}
        \end{array}\right\rbrace
    \quad\quad\quad
    \frac{\partial^2 N}{\partial x_u^2} =
        \sum_{(u,v) \in E} \eta_{\sigma}(x_u-x_v)
\]
and similarly in the $y$-dimension.

For grid-snap forces we have
\[
    \frac{\partial G}{\partial x_u} =
        \gamma_{\sigma}(x_u-a_u)
\]
and
\[
    \frac{\partial^2 G}{\partial x_v \partial x_u} = 0
    \quad\quad\quad
    \frac{\partial^2 G}{\partial x_u^2} = \eta_{\sigma}(x_u-a_u)
\]
and similarly in the $y$-dimension.
Recall that $(a_u,b_u)$ is defined to be the closest grid point to $(x_u,y_u)$.

For edge-node repulsion forces we have
\[
    \frac{\partial E}{\partial x_u} =
        \sum_{e \in E_V}
        \mathrm{sgn}(x_u-x_e)
        \gamma_{\sigma}\left( (\sigma-d(u,e))^+ \right)
\]
where $x_e$ is the $x$-coordinate of a vertically aligned edge $e$,
and
\[
    \frac{\partial^2 E}{\partial x_v \partial x_u} = 0
    \quad\quad\quad
    \frac{\partial^2 E}{\partial x_u^2} =
        \sum_{e \in E_V}
        \eta_{\sigma}(\sigma-d(u,e))
\]
and similarly in the $y$-dimension.

\subsection{Proof of Theorem 1}
%
%

We begin with definitions and notation.

\medskip
\noindent {\bf Definition:}
A \emph{constrained graph} is an ordered triple $G = (V,E,C)$ where
$V$ and $E$ are the sets of nodes and edges in the graph, and $C$
is a set of separation constraints on the $x$- and $y$-coordinates of
the nodes.

\medskip
\noindent {\bf Notation:}
The results of this section hold for both directed and undirected
graphs. Since the directedness of edges is completely irrelevant to
our results, we write an edge whose endpoints are $u$ and $v$ in
unordered notation $\{u,v\}$.

\medskip
\noindent {\bf Definition:}
An \emph{edge constraint} for a graph $G=(V,E)$ is a
separation constraint $z_u + g \leq z_v$ or $z_u + g = z_v$
such that $\{u,v\} \in E$, i.e.~a separation constraint on the endpoints
of an edge.

\medskip
\noindent {\bf Definition:}
An \emph{edge-constrained graph} is a constrained graph $G = (V,E,C)$
in which $C$ contains only edge constraints.

\medskip
\noindent {\bf Notation:}
When $C$ is a set of constraints and $S$ a set of equations and
inequalities on coordinates of nodes, we will use the entailment
relation $C \vdash S$ to indicate that each relation in $S$ is
entailed by the constraints in $C$.

\medskip
\noindent {\bf Definition:}
A constrained graph $G = (V,E,C)$ is said to contain a
\emph{horizontal overlay} when there are edges
$\{a,b\}, \{c,d\} \in E$ such that
\[
  C \vdash \{y_a = y_b = y_c = y_d, x_a < x_d, x_c < x_b\}.
\]
In this case we write $(a,b) \hlay (c,d)$.
Similarly, $G$ is said to contain a \emph{vertical overlay} when there
are edges $\{e,f\}, \{g,h\} \in E$ such that
\[
  C \vdash \{x_e = x_f = x_g = x_h, y_e < y_h, y_g < y_f\},
\]
in which case we write $(e,f) \vlay (g,h)$.

\medskip
\noindent {\bf NB:} While edges are written in undirected notation,
the overlay notation \emph{is ordred}. For example,
$(a,b) \hlay (c,d)$ is different from $(b,a) \hlay (c,d)$.

\medskip
Since the horizontal and vertical cases of Theorem~1 are entirely
similar, we prove only the horizontal case. We begin with a
lemma.

\medskip
\noindent {\bf Lemma:}
If an edge-constrained graph $G = (V,E,C)$ contains a horizontal overlay,
then there exist three nodes $u, v, w \in V$ such
that $\{u,v\}, \{u,w\} \in E$ and either
$(u,v) \hlay (u,w)$ or $(v,u) \hlay (w,u)$.

\medskip
\noindent {\bf Proof:}
By the definition of horizontal overlay there are edges
$\{a,b\}, \{c,d\} \in E$ such that $C \vdash (a,b) \hlay (c,d)$.
Let $H = (V,F)$ be the graph in which $\{e,f\} \in F$
if and only if $\{e,f\} \in E$ and $C \vdash y_e = y_f$.
Let $K$ be the connected component of $a$ in $H$.
Since $G$ is edge-constrained, we have $a, b, c, d \in K$.
Note that all nodes in $K$ share one and the same $y$-coordinate.
We will find $u, v, w \in K$ satisfying the statement of
the lemma.

To begin with, let $deg_H$ denote degree in $H$, and suppose that
there is any $u \in K$ with $deg_H(u) \geq 3$.
Then by the pigeonhole principle $u$ must either have two neighbours
$v, w \in K$ on its left, or two on its right, and in either case
we are done.

Suppose then that all $u \in K$ have $1 \leq  deg_H(u) \leq 2$.
Then $K$ forms either a cycle or a chain.
Consider first the case in which $K$ forms a cycle, that is,
every $u \in K$ has $deg_H(u) = 2$.
Then if $u, v, w \in K$ could not be found to satisfy the lemma,
then for all $v \in K$ we would have $x_u < x_v < x_w$, where
$u$ and $w$ are the two neighbours of $v$. In this case we would
have a cycle of less-than relations, making $x_u < x_u$ for each
$u \in K$, which is impossible.

This leaves only the case in which $K$ forms a chain.
Again, if $u, v, w \in K$ could not be found to satisfy the lemma,
then each $v \in K$ with $deg_H(v) = 2$ would have one of its
neighbours on each side of it, so that $K$ would contain no overlay
at all, contrary to assumption. This proves the lemma.

\medskip
Finally we restate Theorem 1 for the case of horizontal overlays in terms
of the definitions of this section, and prove it.

\medskip
\noindent {\bf Theorem~1:} Let $G=(V,E,C)$ be an edge-constrained graph with
$\{u,v\} \in E$, having no constraints relating $u$ and $v$, and
containing no horizontal overlays.
Let $S = \mathsf{SA}(u,v,\mathbb{E})$.
Then $C \cup \{S\}$ entails a horizontal overlay if and only if
there exists a node $w \in V$
such that $C \vdash y_w = y_u$ or $C \vdash y_w = y_v$, and
satisfying one of the following
two sets of conditions:
\begin{enumerate}
\item
    \begin{enumerate}
    \item $\{u,w\} \in E$, and
    \item $C \vdash x_u < x_w$ or $C \vdash x_v < x_w$, or
    \end{enumerate}
\item
    \begin{enumerate}
    \item $\{w,v\} \in E$, and
    \item $C \vdash x_w < x_u$ or $C \vdash x_w < x_v$.
    \end{enumerate}
\end{enumerate}

\medskip
\noindent {\bf Proof:}
It is clear that if a node $w$ satisfying the stated conditions
exists, then a horizontal overlay will be created
when $S$ is applied.
Conversely, we now suppose that a horizontal overlay is created when $S$ is
applied, and prove that such a node $w$ must exist.

Let $C' = C \cup \{S\}$. By the Lemma there exist three nodes
$a, b, c \in V$ with $\{a,b\}, \{a,c\} \in E$ such that
$C' \vdash (a,b) \hlay (a,c)$ or
$C' \vdash (b,a) \hlay (c,a)$.
But since neither of these overlays is entailed by $C$, we can conclude that
one of the edges $\{a,b\}, \{a,c\}$ has to be $\{u,v\}$.
We assume (renaming if necessary) that $\{a,b\} = \{u,v\}$,
and show that taking $w = c$ satisfies the conditions of the theorem.
Specifically, we will handle the case in which $a = u$.
The case in which $a = v$ is similar.

In this case $C' \vdash (v,u) \hlay (c,u)$ cannot occur, since
this would involve the entailment $C' \vdash x_v < x_u$, whereas
we assumed that $C$ states no relation on nodes $u$ and $v$, while
the only order relation entailed by $S$ is $x_u < x_v$.
Therefore we must have
$C' \vdash (u,v) \hlay (u,c)$, which says that
\[
    C' \vdash \{ y_u = y_v = y_c, x_u < x_v, x_u < x_c\}.
\]
Since $y_u = y_v$ is the only equation entailed by $S$, we
conclude that $C \vdash y_c = y_u$ or $C \vdash y_c = y_v$.
By assumption, $\{u,c\} \in E$.
And again, since the only inequality entailed by $S$
relates $x_u$ and $x_v$, it must be that
$C \vdash x_u < x_c$ or $C \vdash x_v < x_c$.
This completes the proof.

\section{Detailed Results}
\begin{figure}
\centering
\includegraphics[width=\textwidth]{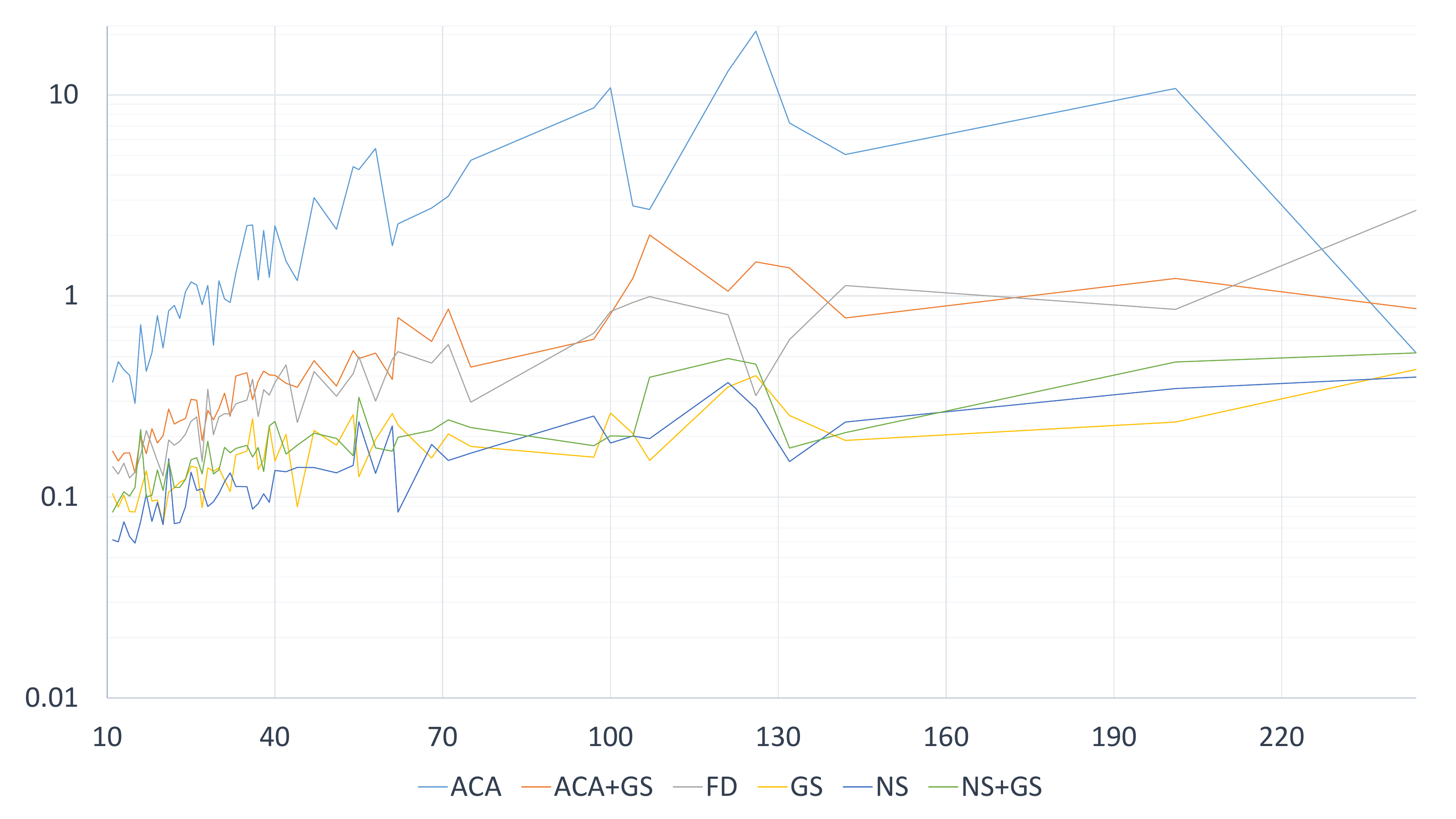}
\caption{Running time in seconds for the six different grid-like layout methods against number of nodes for the 252 graphs in our corpus.  Times given do not include the other layout stages.  For example, ACA does not include the initial FD layout.  ACA+GS, is just the additional grid stage after FD+ACA.}
\label{fig:eval-time}
\end{figure}

\begin{figure}
\centering
\includegraphics[width=\textwidth]{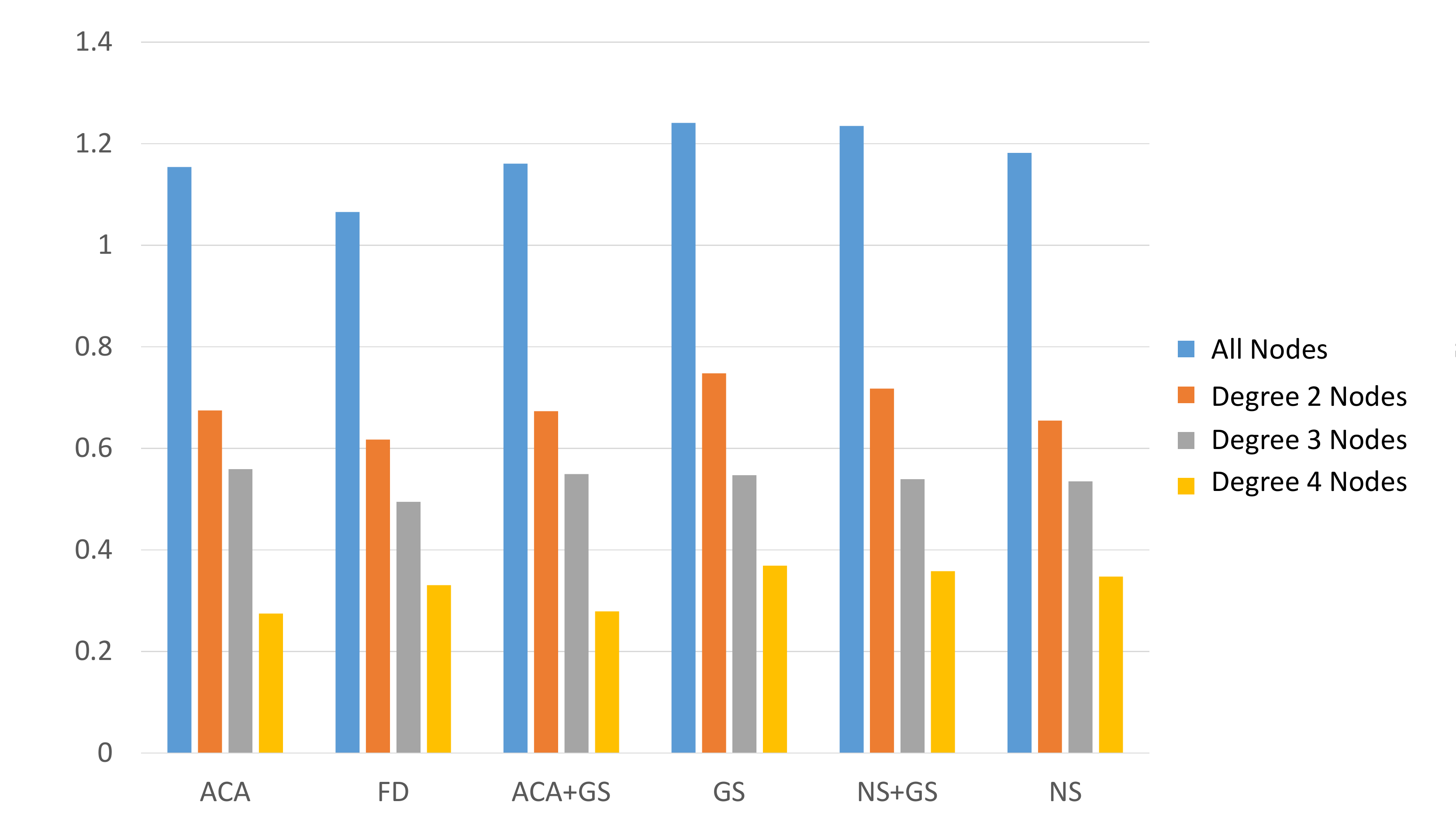}
\caption{Angular resolution for the various techniques for all nodes,
but also broken down for lower degree nodes.
We see ACA does almost as well as FD on degree-2 nodes, and
results in better angular resolution than FD for degree-4.
This is expected since---as explained in Section 4---the heuristic does target these.}
\label{fig:eval-angres}
\end{figure}

\begin{figure}
\centering
\includegraphics[width=\textwidth]{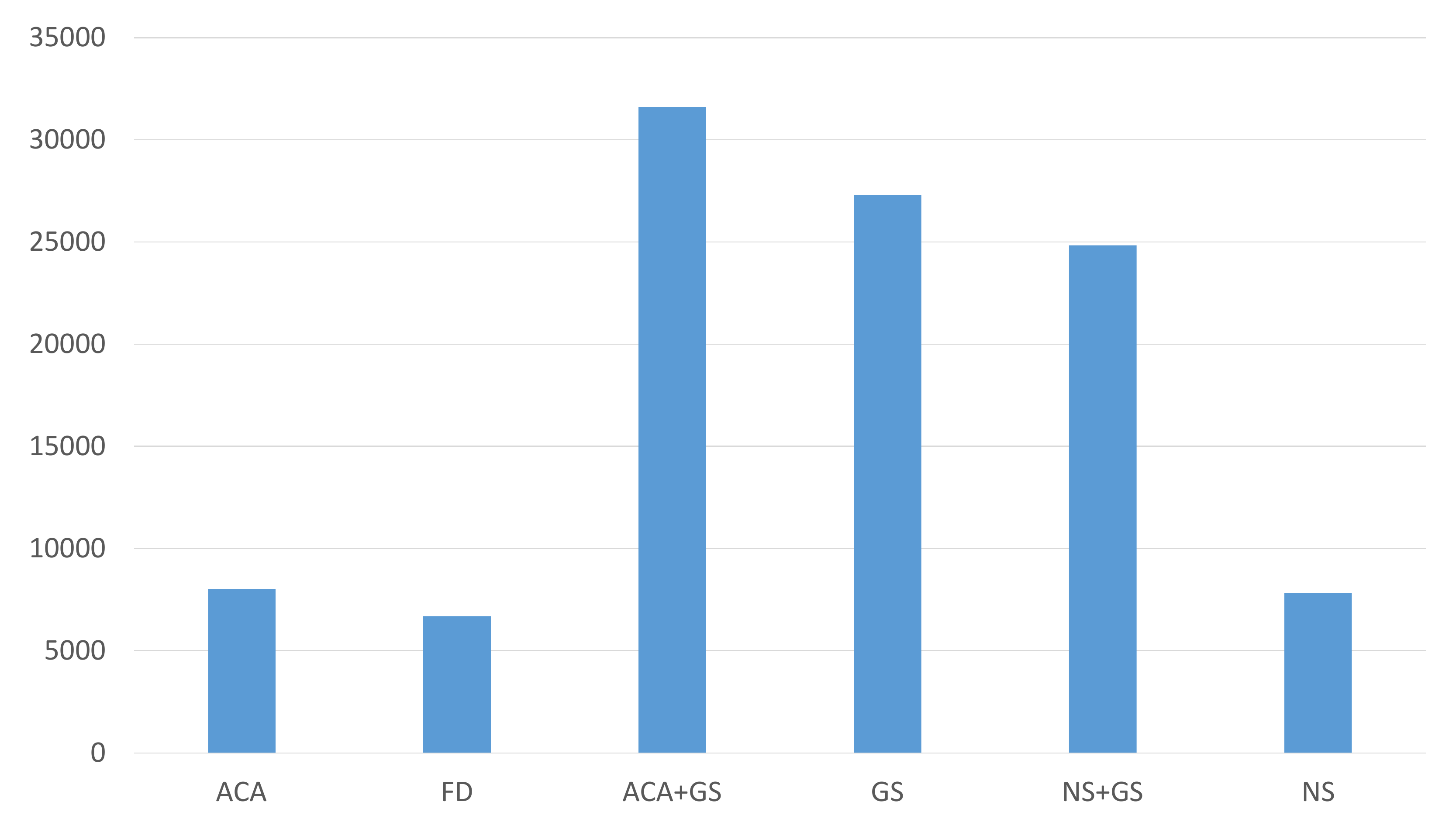}
\caption{P-Stress values normalized by graph size and density.  There is not much to note except that GS introduces the most significant stress.  Basically, this means that optimization over
\emph{GS-stress} introduces the most distortion of the underlying FD layout.}
\label{fig:eval-stress}
\end{figure}

\begin{figure}
\centering
\subfigure[FD]{
\includegraphics[width=0.3\textwidth]{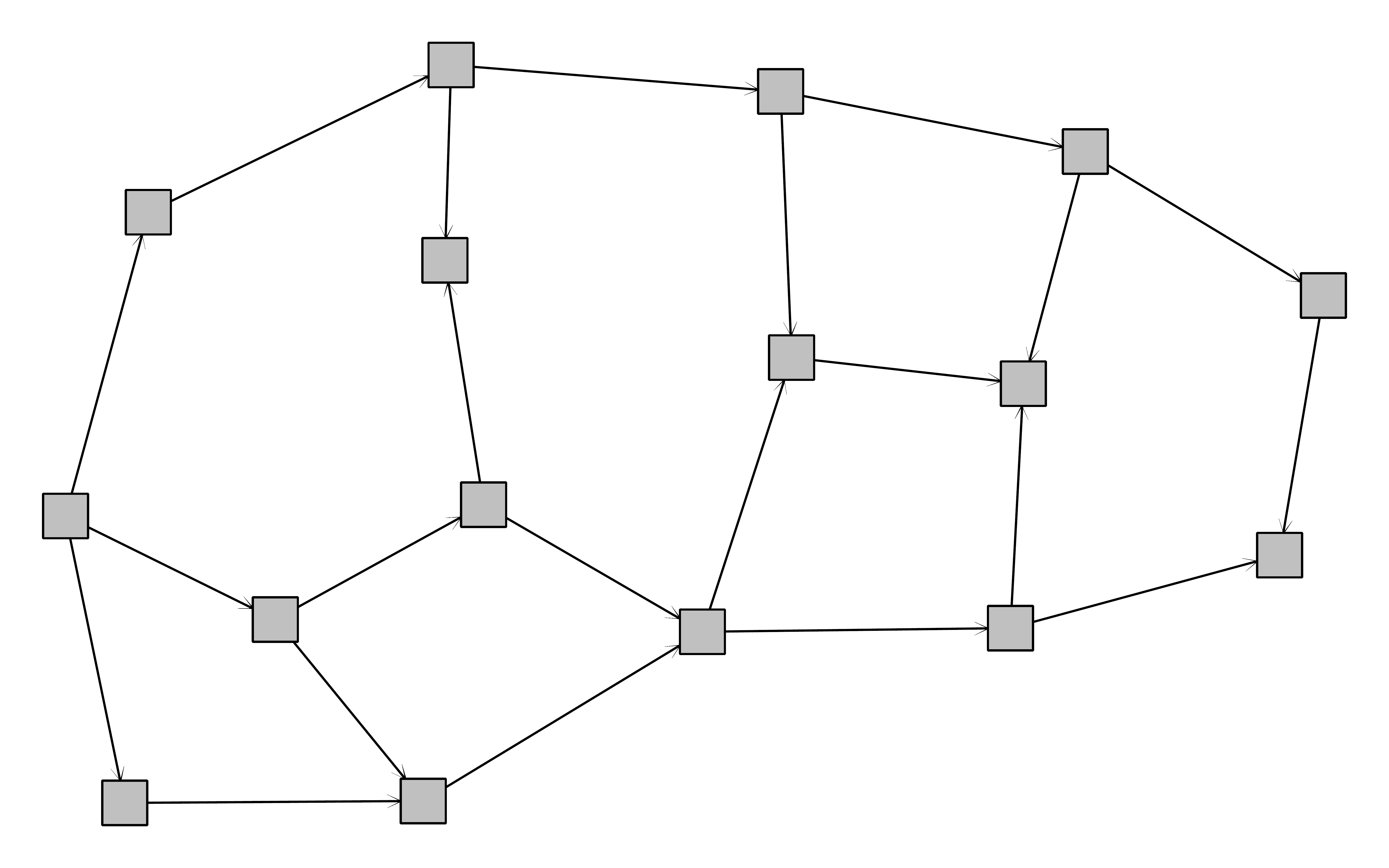}
}
\subfigure[GS]{
\includegraphics[width=0.3\textwidth]{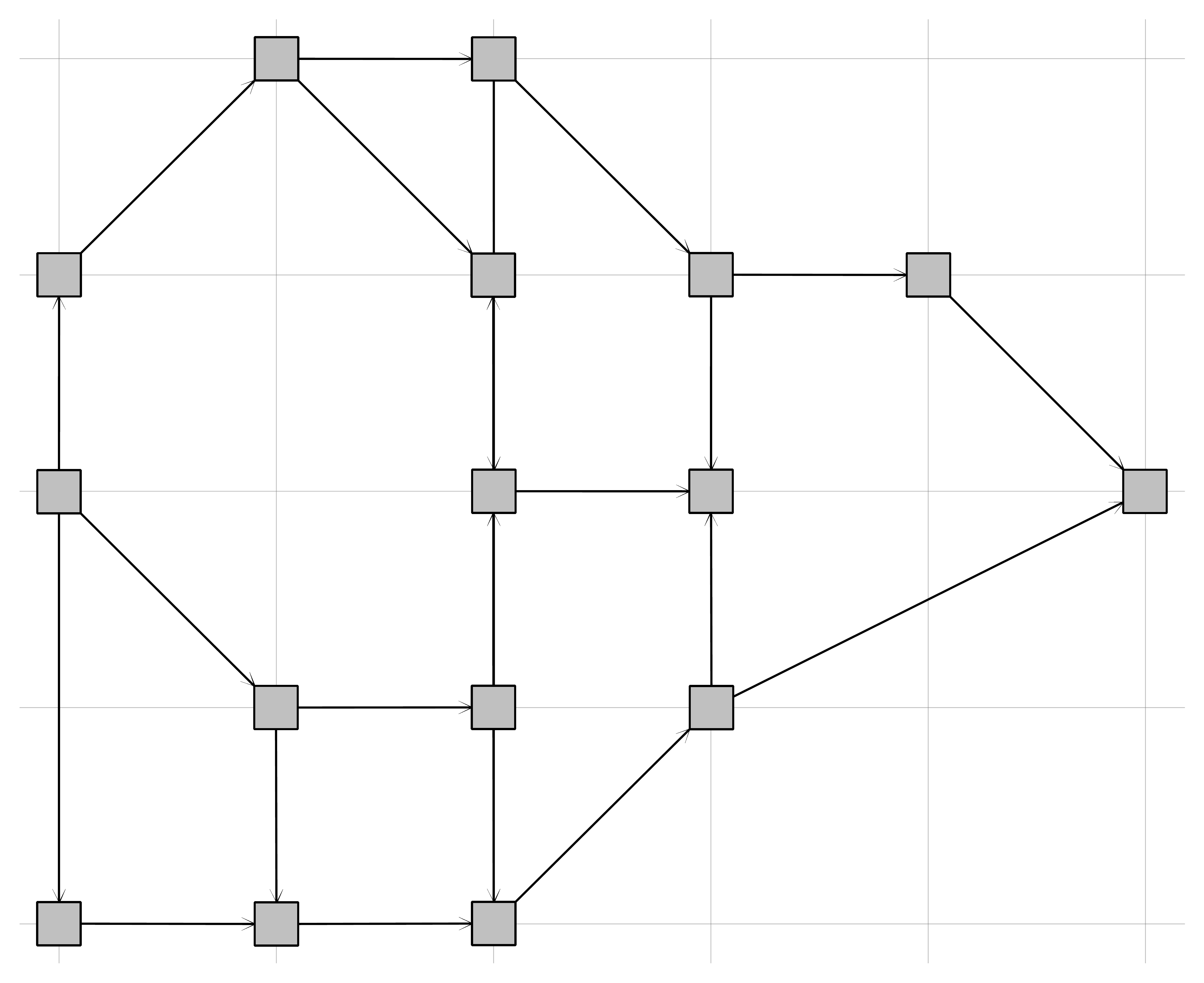}
}
\subfigure[NS]{
\includegraphics[width=0.3\textwidth]{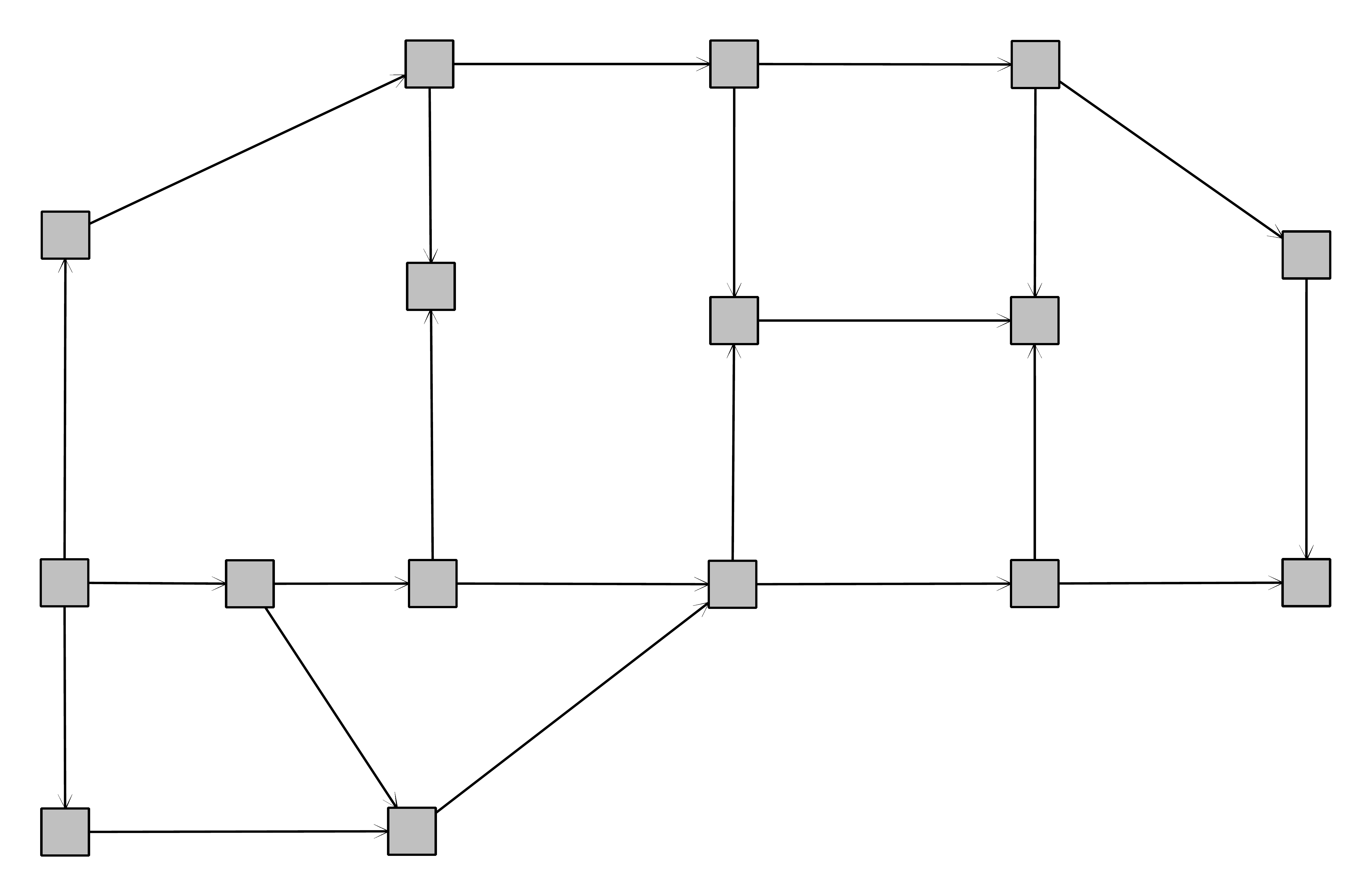}
}
\subfigure[NS + GS]{
\includegraphics[width=0.3\textwidth]{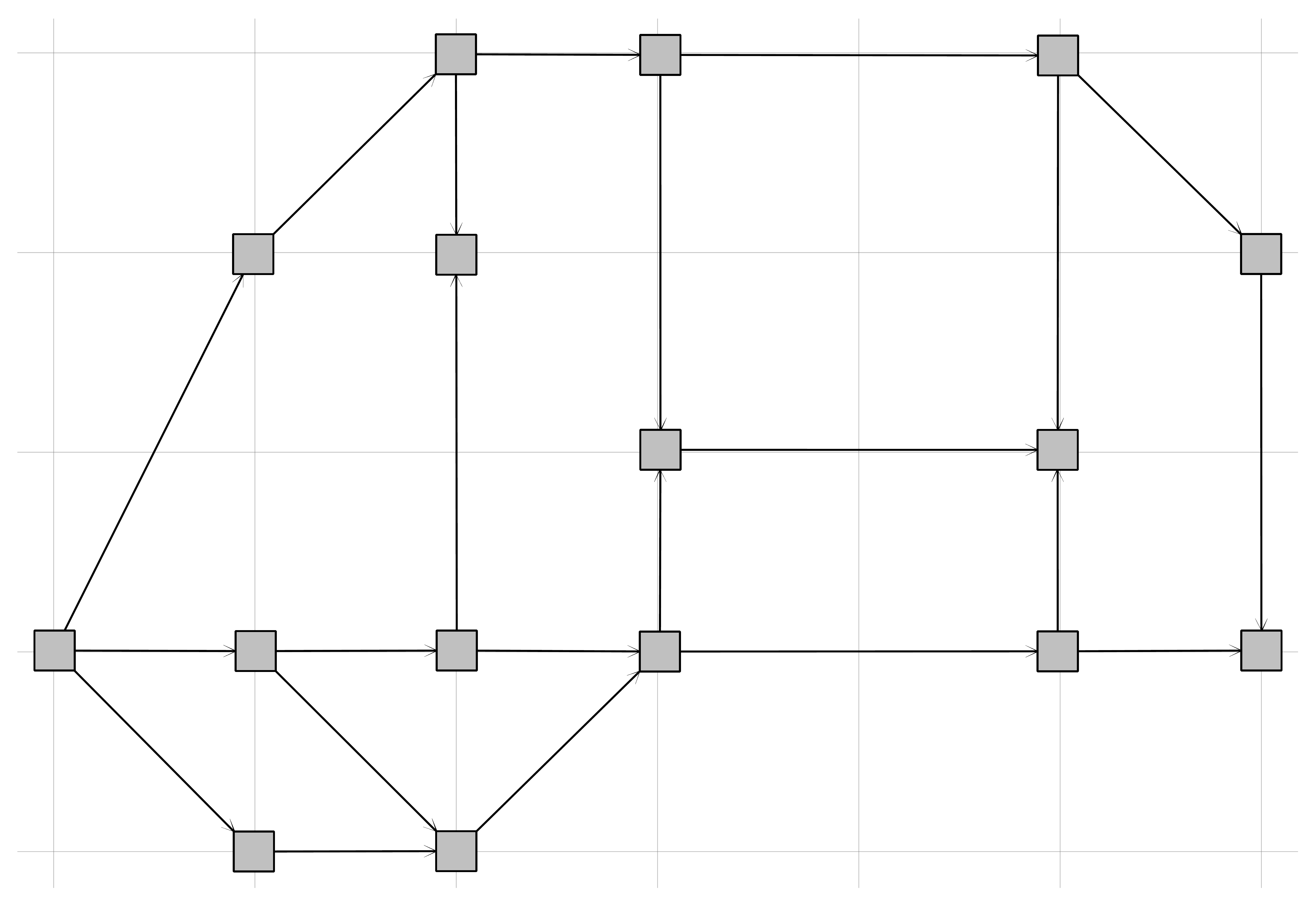}
}
\subfigure[ACA]{
\includegraphics[width=0.3\textwidth,trim=-4cm 0cm -4cm 0cm]{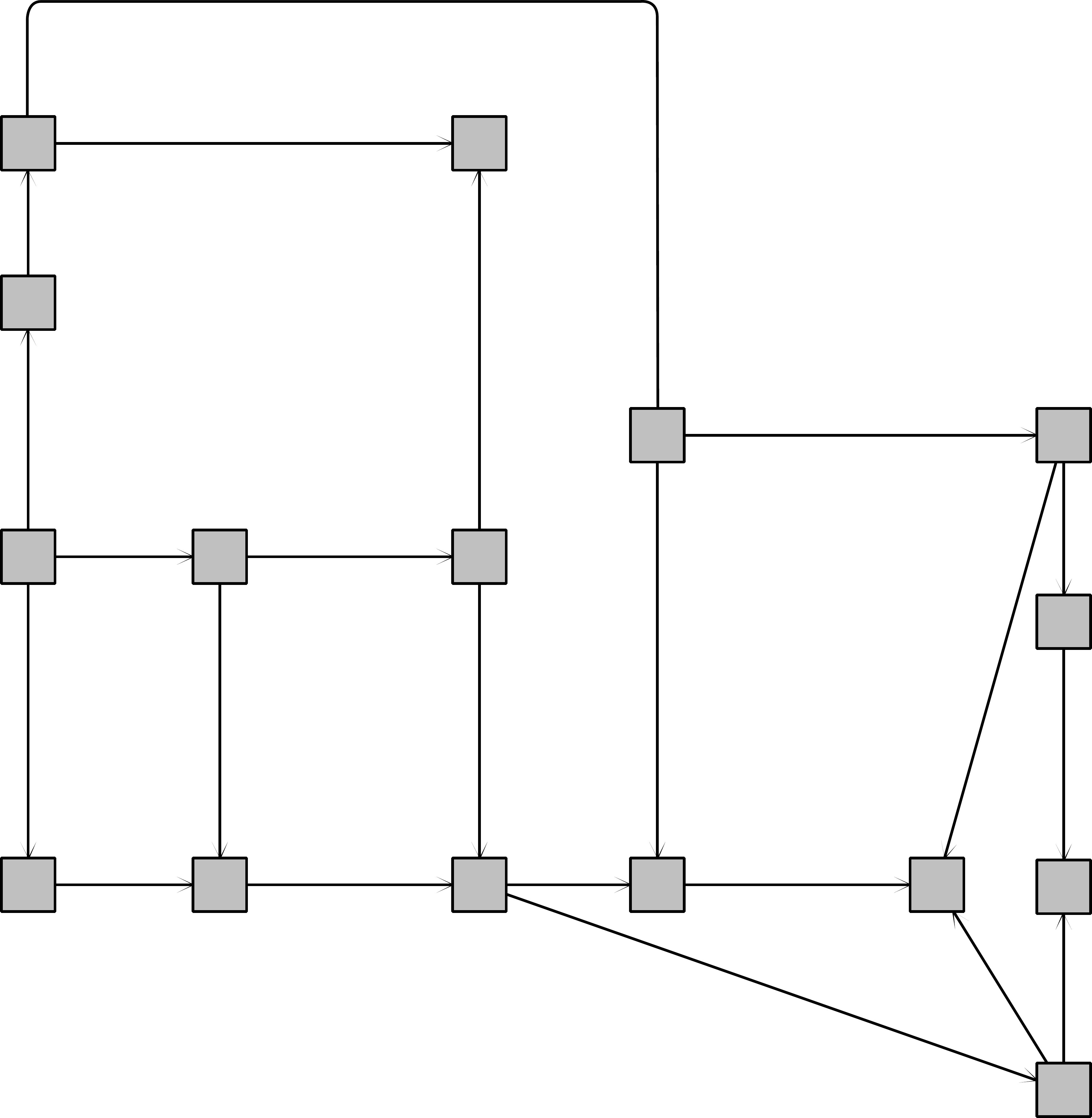}
\label{subfig:routed1}
}
\subfigure[ACA + GS]{
\includegraphics[width=0.29\textwidth]{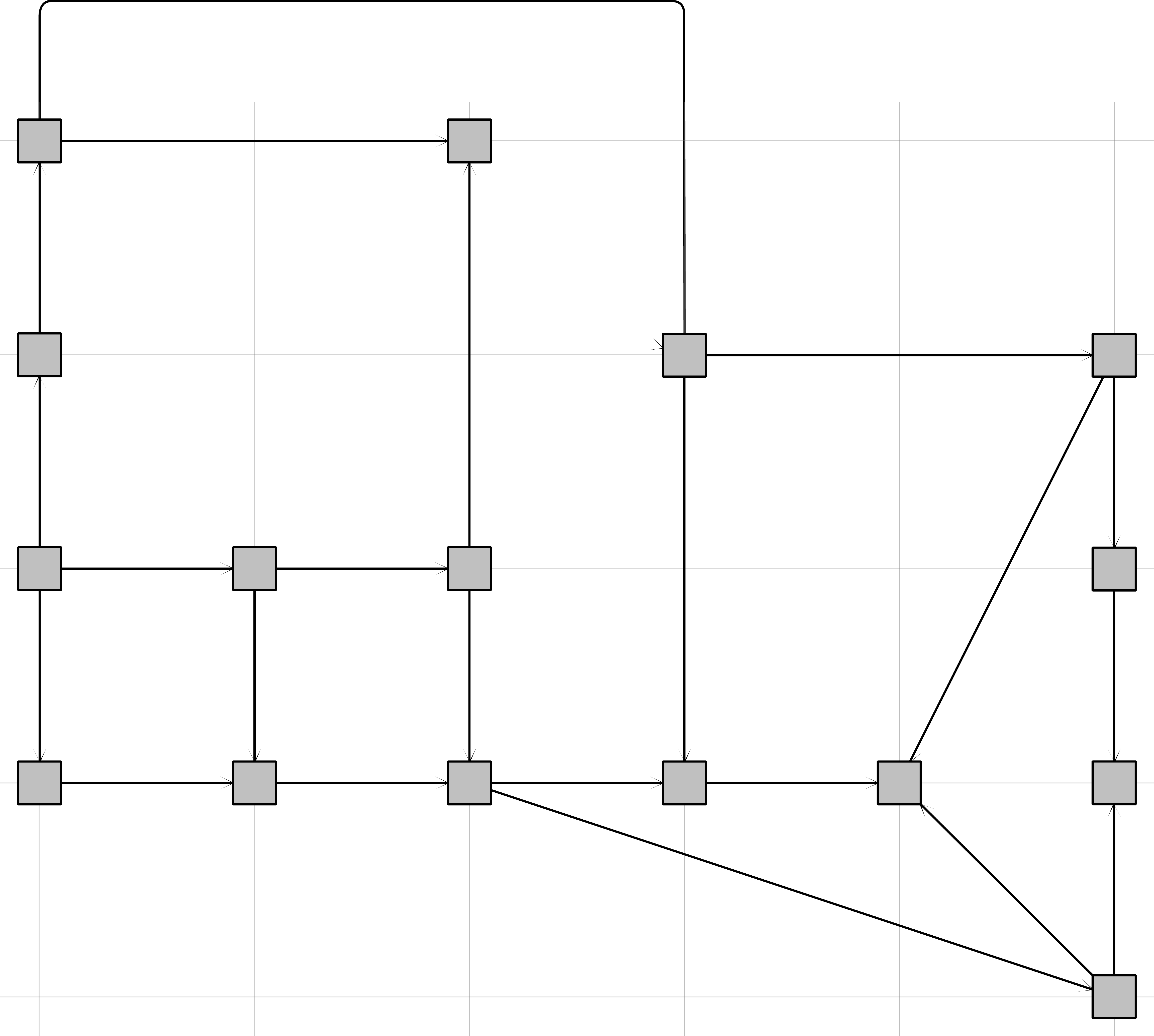}
\label{subfig:routed2}
}
\caption{Different combinations of our automatic layout techniques for the graph ``ug\_213'' from the AT\&T Graphs corpus, as generated during our evaluation.  In \ref{subfig:routed1} and \ref{subfig:routed2} we use a simple post-process to see if edges involved in crossings can be rerouted to avoid crossings using the orthogonal connector routing scheme described in \cite{wybrow2010orthogonal}.}
\label{fig:eval-213}
\end{figure}

\begin{figure}
\centering
\subfigure[FD]{
\includegraphics[width=0.3\textwidth]{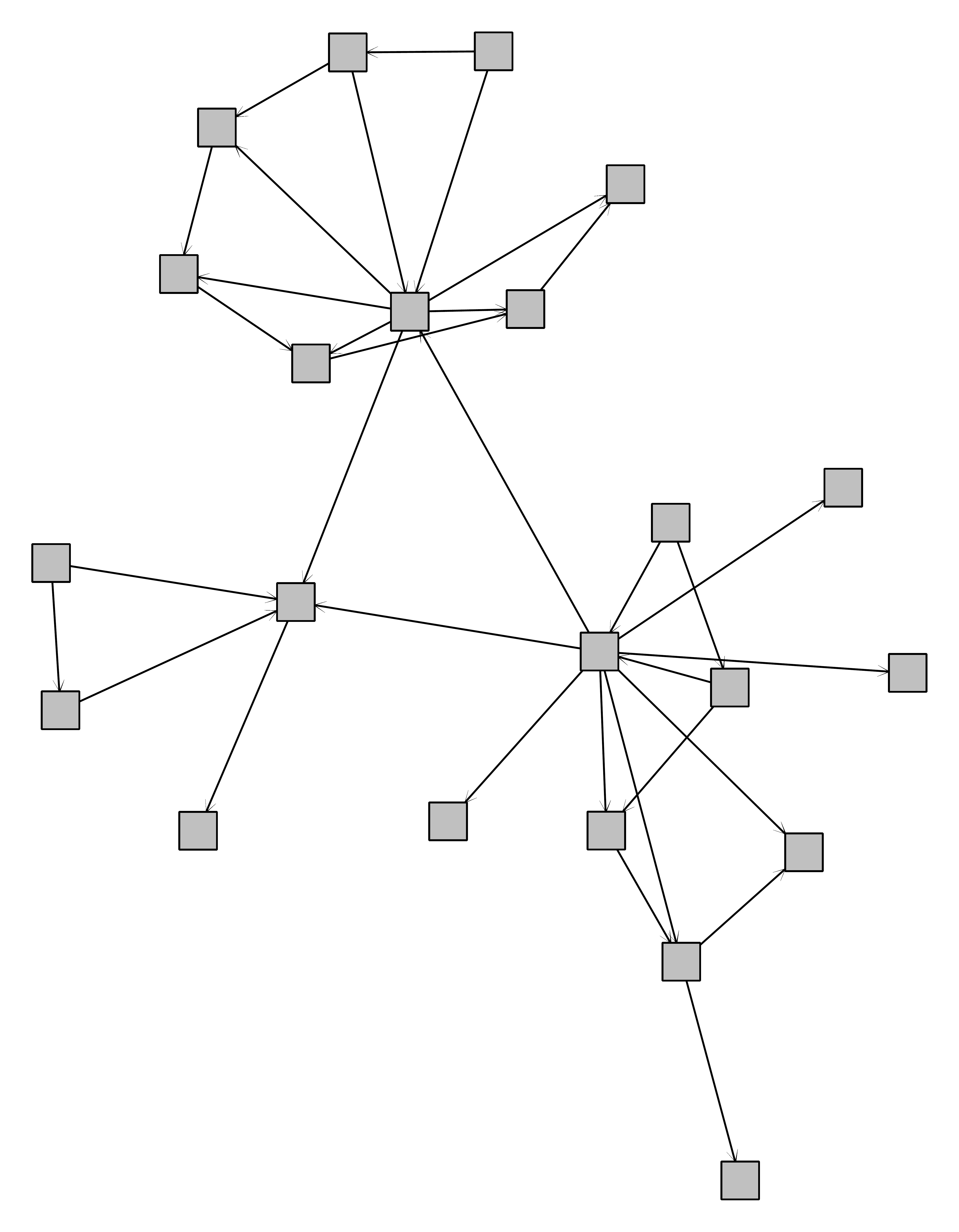}
}
\subfigure[GS]{
\includegraphics[width=0.3\textwidth]{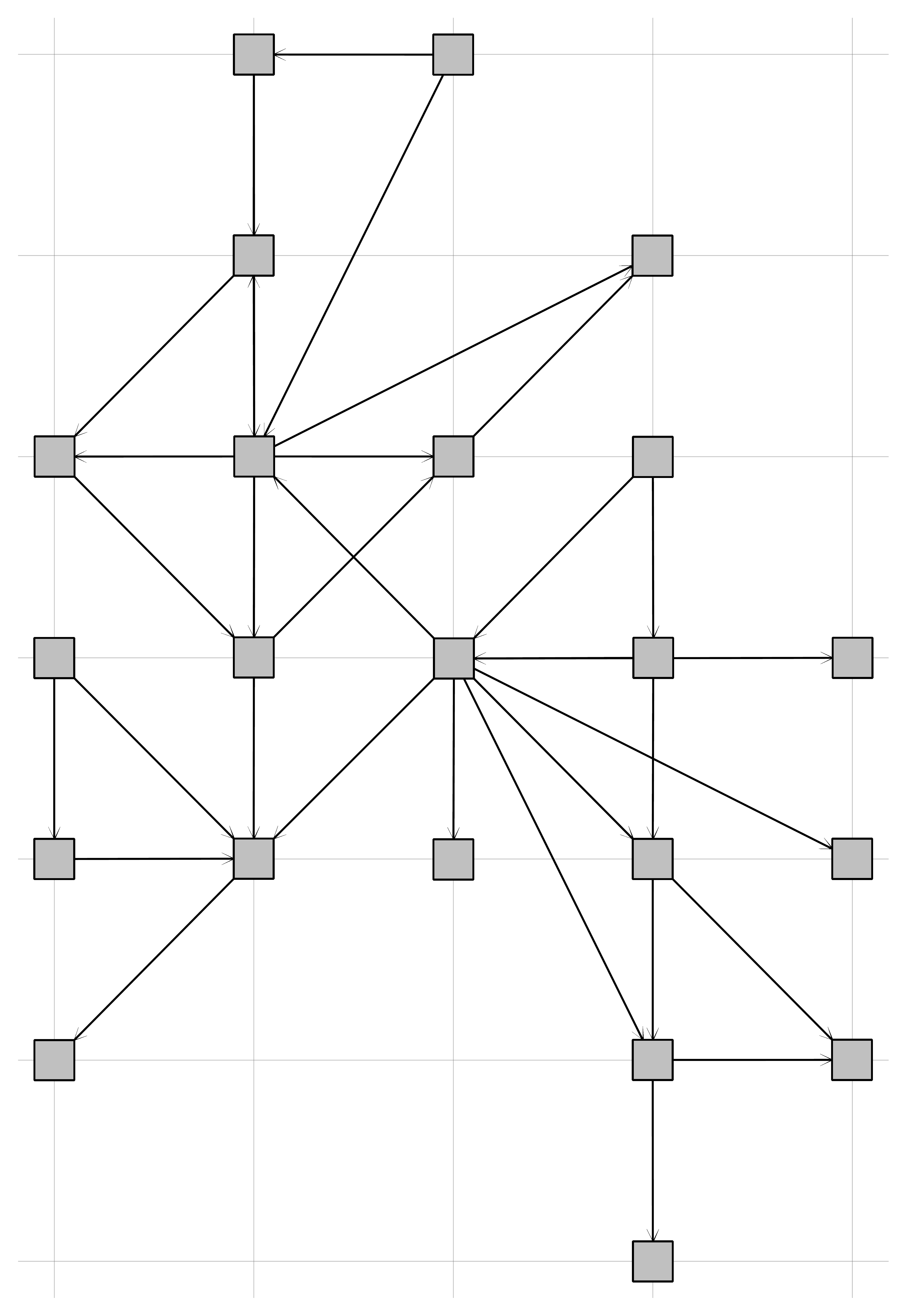}
}
\subfigure[NS]{
\includegraphics[width=0.3\textwidth]{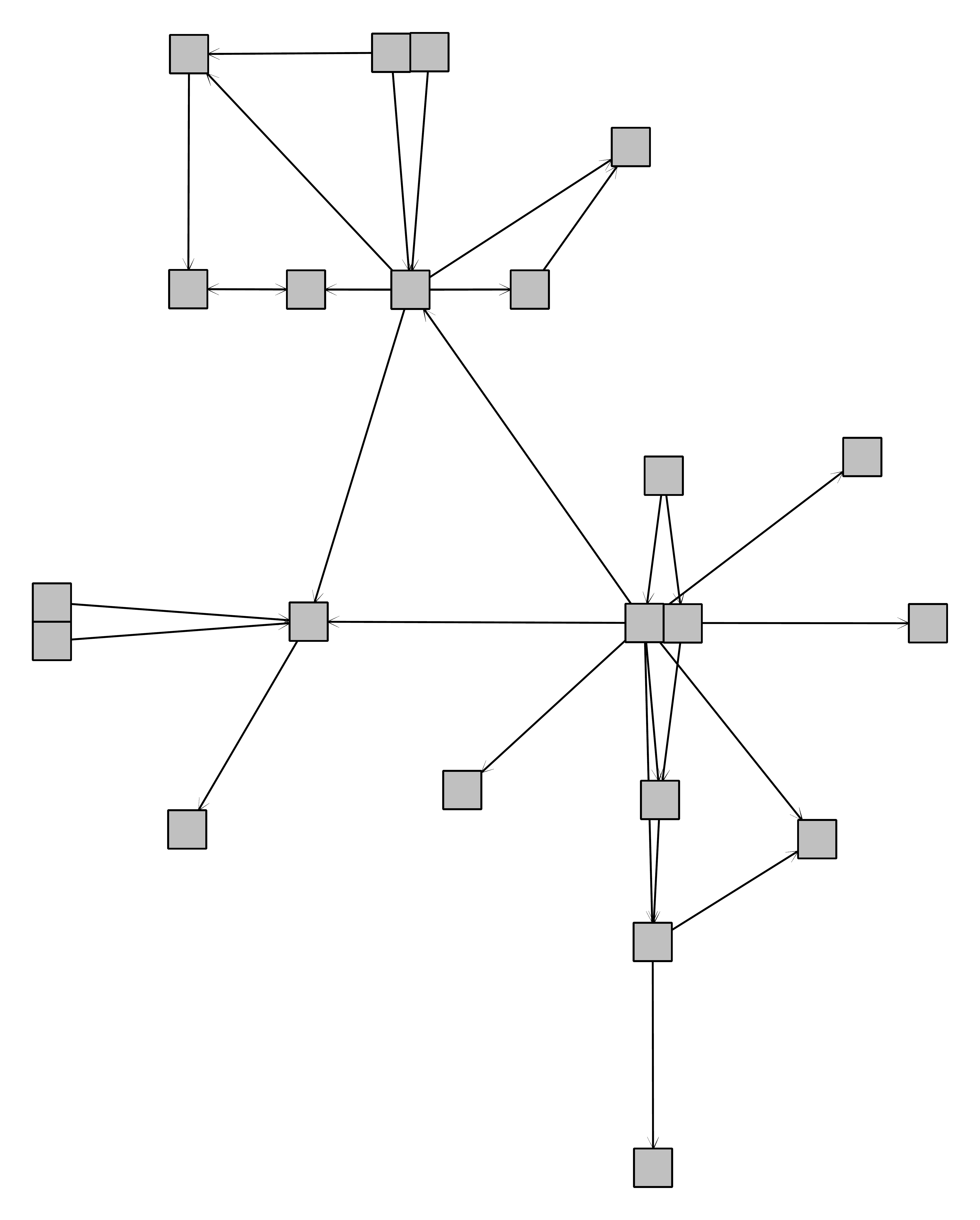}
}
\subfigure[NS + GS]{
\includegraphics[width=0.3\textwidth]{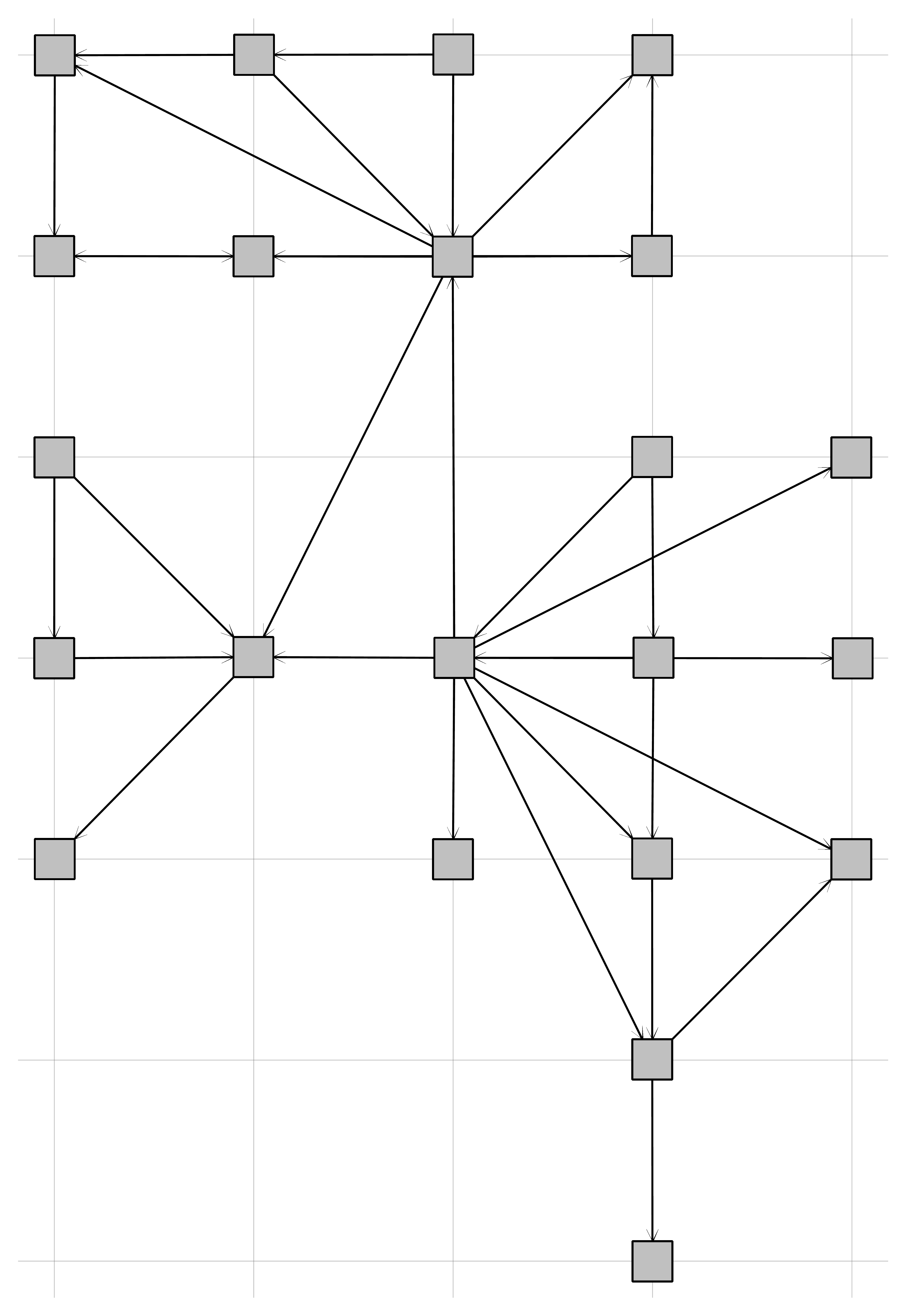}
}
\subfigure[ACA]{
\includegraphics[width=0.3\textwidth]{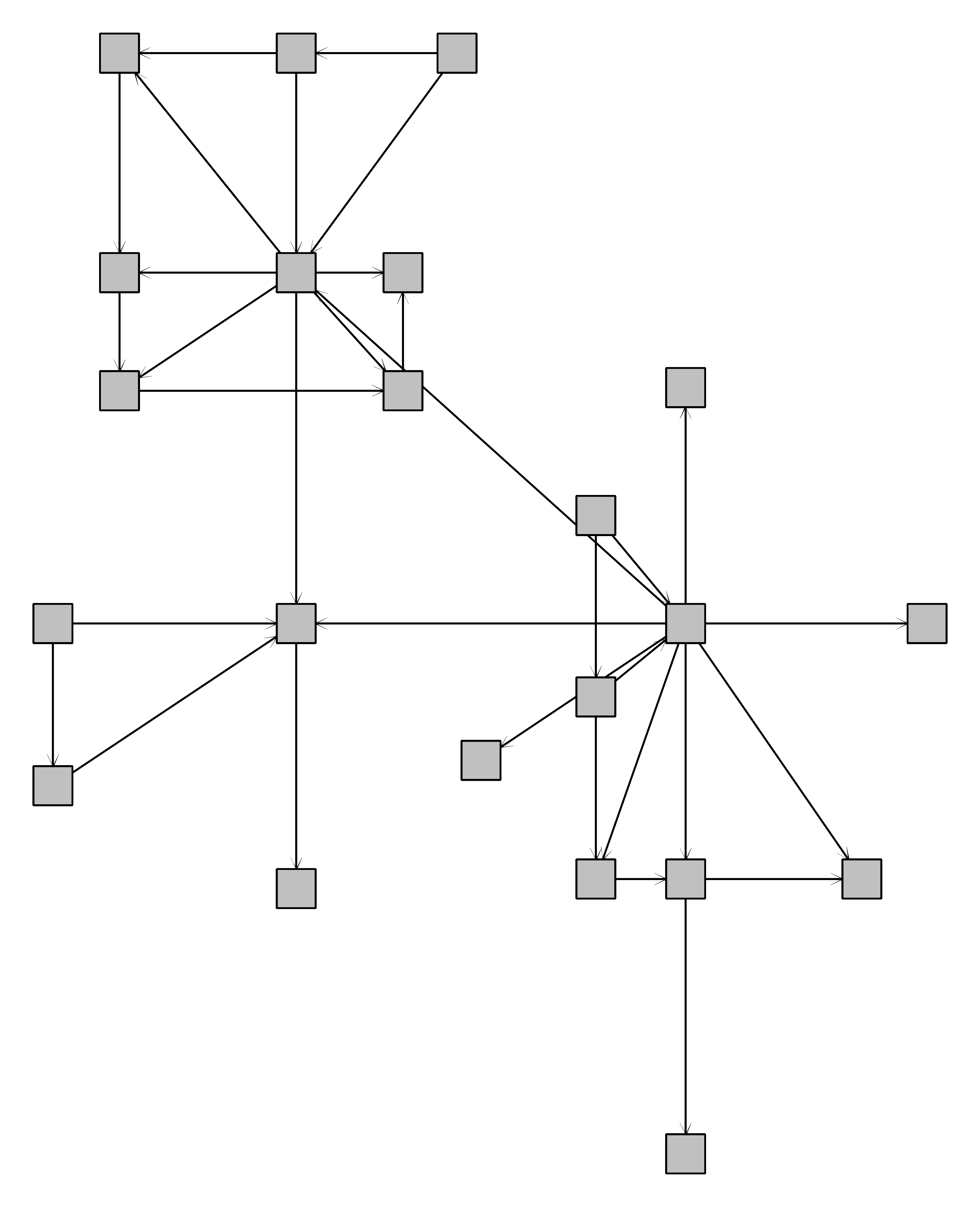}
}
\subfigure[ACA + GS]{
\includegraphics[width=0.3\textwidth]{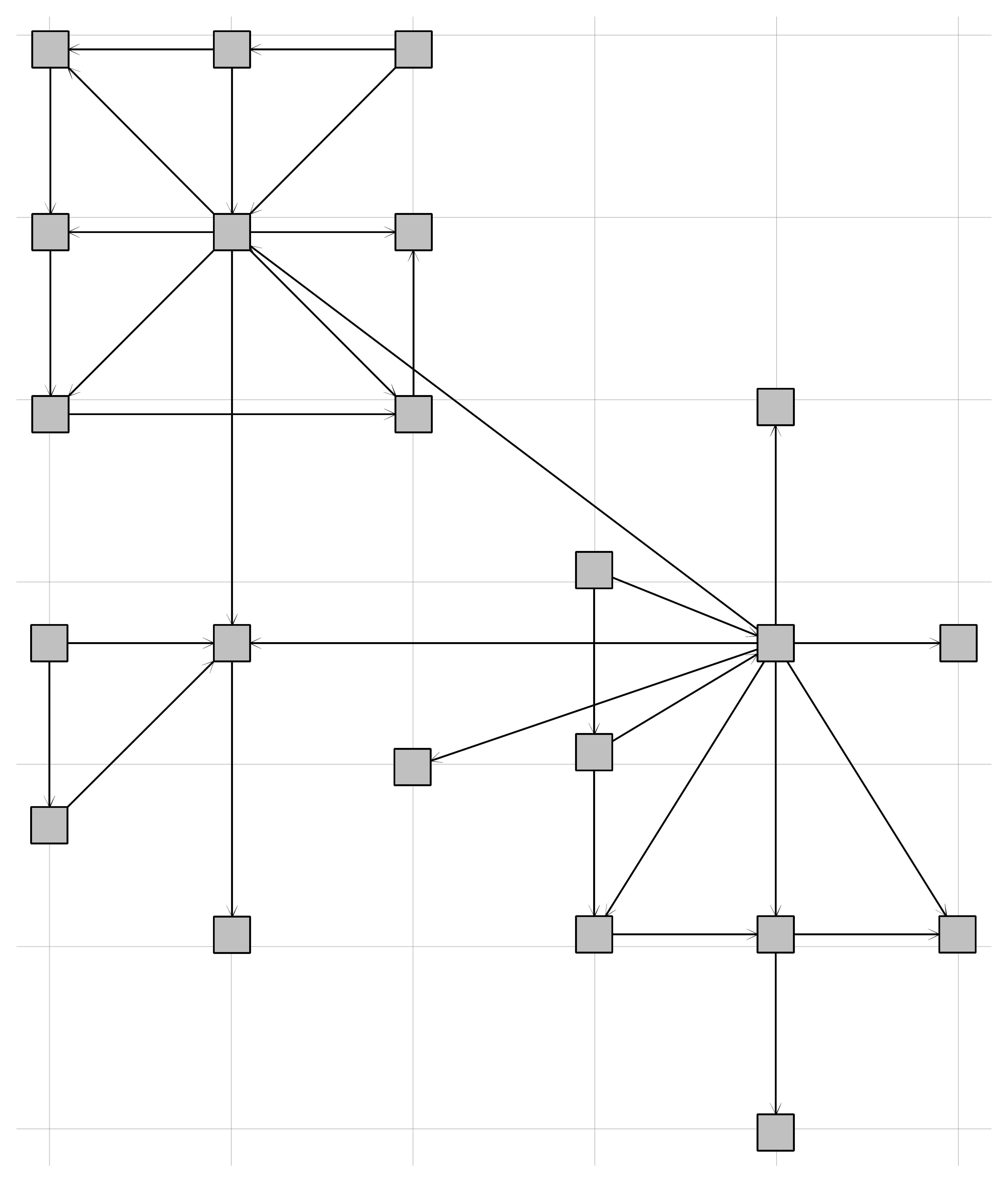}
}
\caption{Different combinations of our automatic layout techniques for the graph ``ug\_268'' from the AT\&T Graphs corpus, as generated during our evaluation.}
\label{fig:eval-268}
\end{figure}

\begin{figure}
\centering
\subfigure[FD]{
\includegraphics[width=0.3\textwidth]{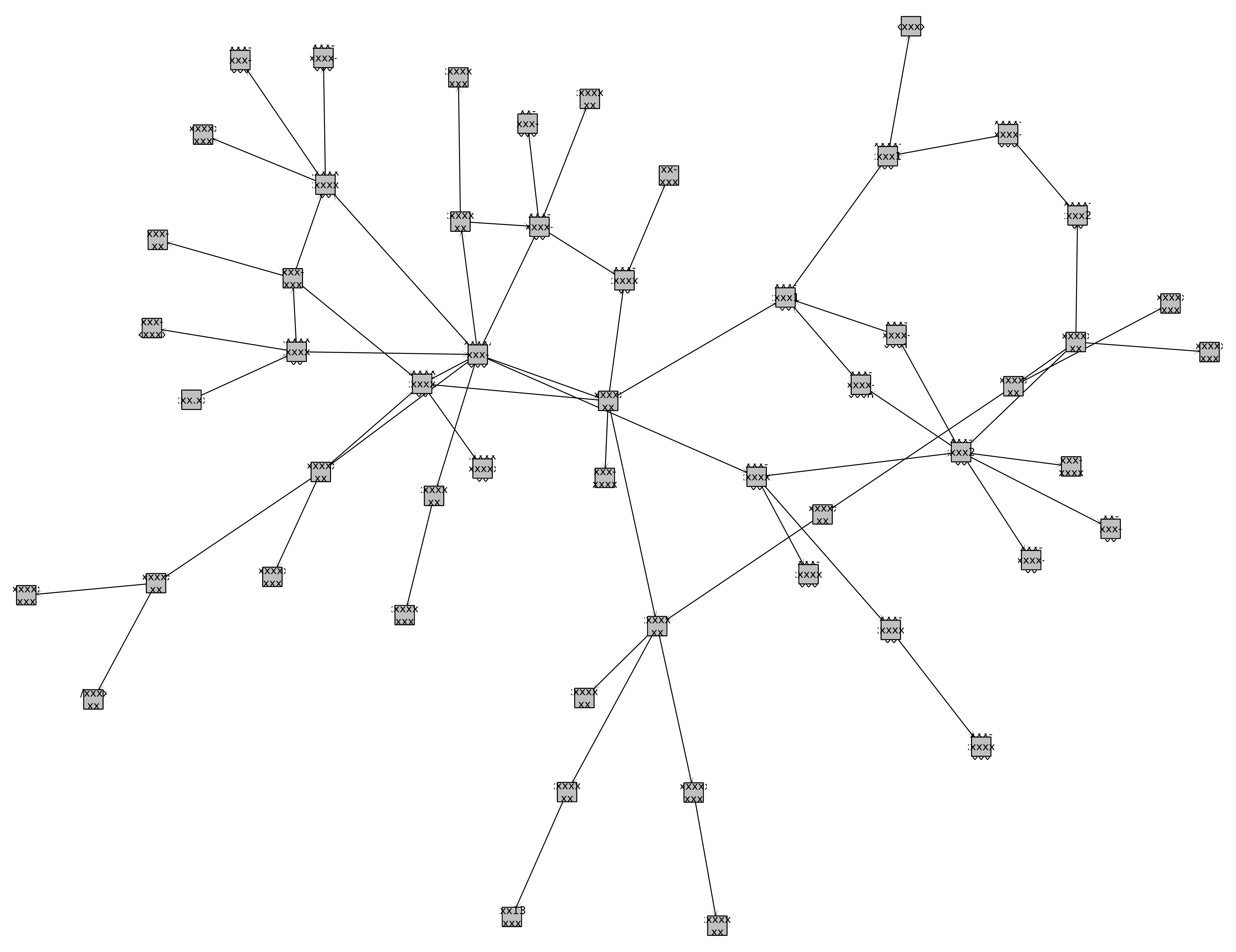}
}
\subfigure[GS]{
\includegraphics[width=0.3\textwidth]{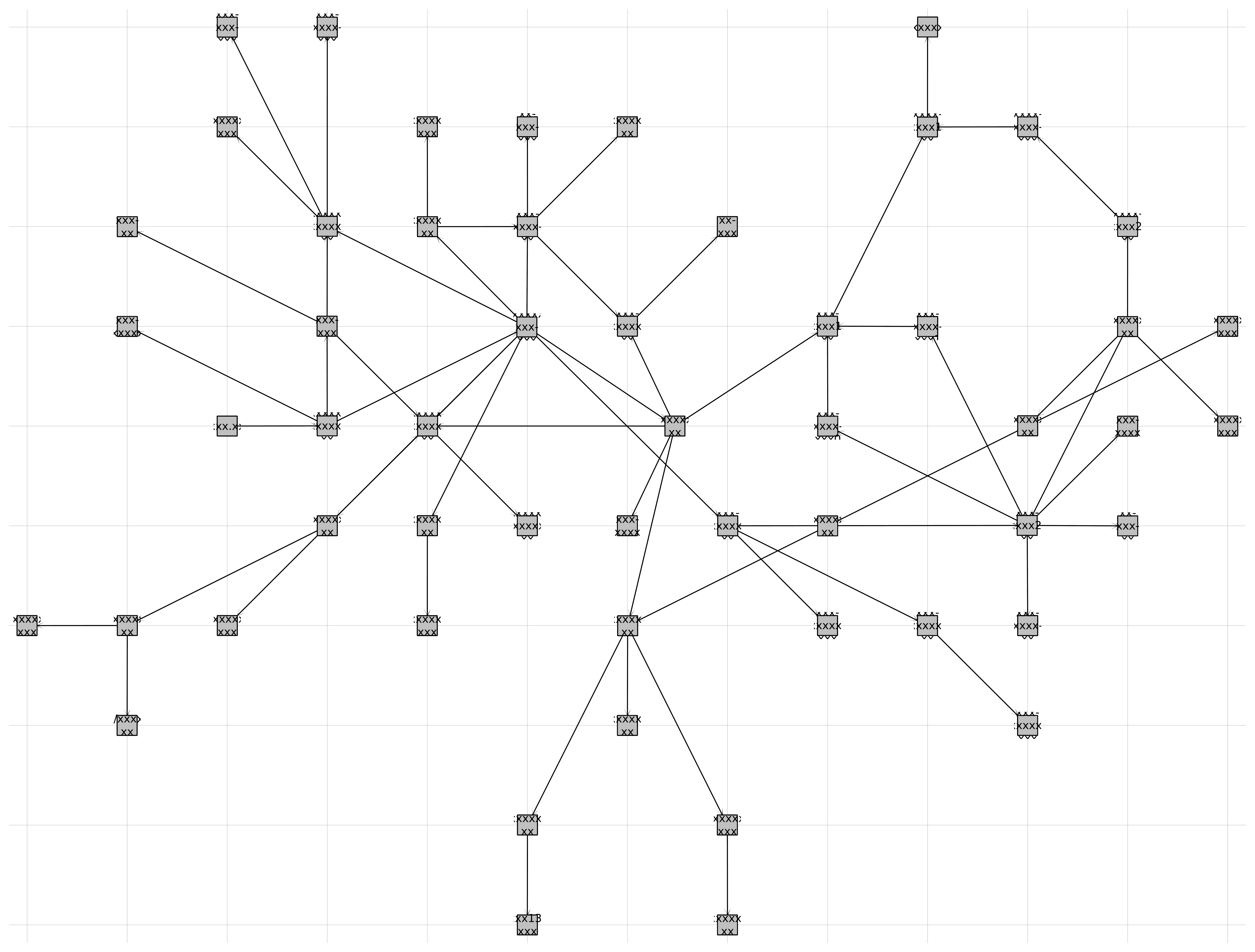}
}
\subfigure[NS]{
\includegraphics[width=0.3\textwidth]{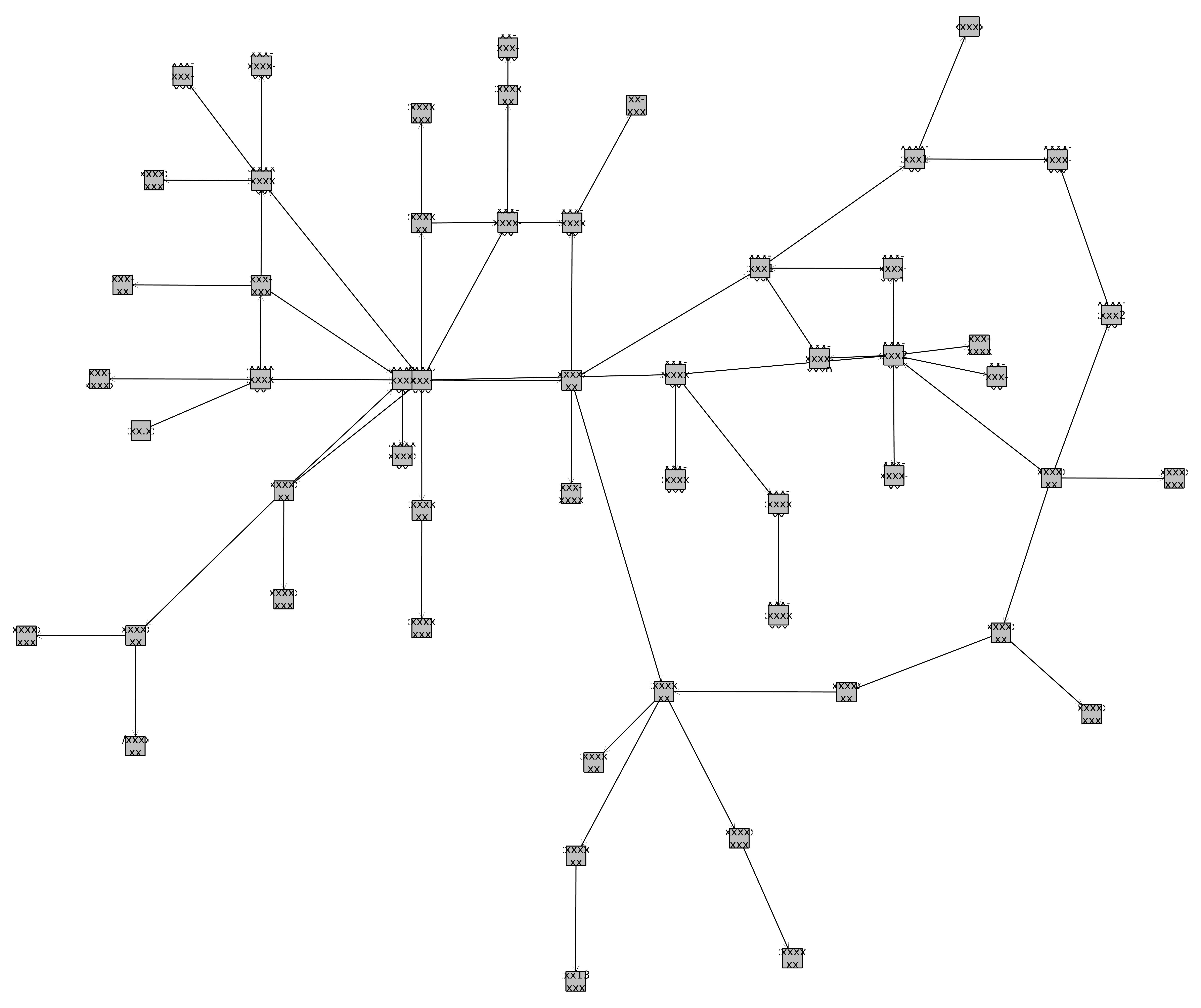}
}
\subfigure[NS + GS]{
\includegraphics[width=0.3\textwidth]{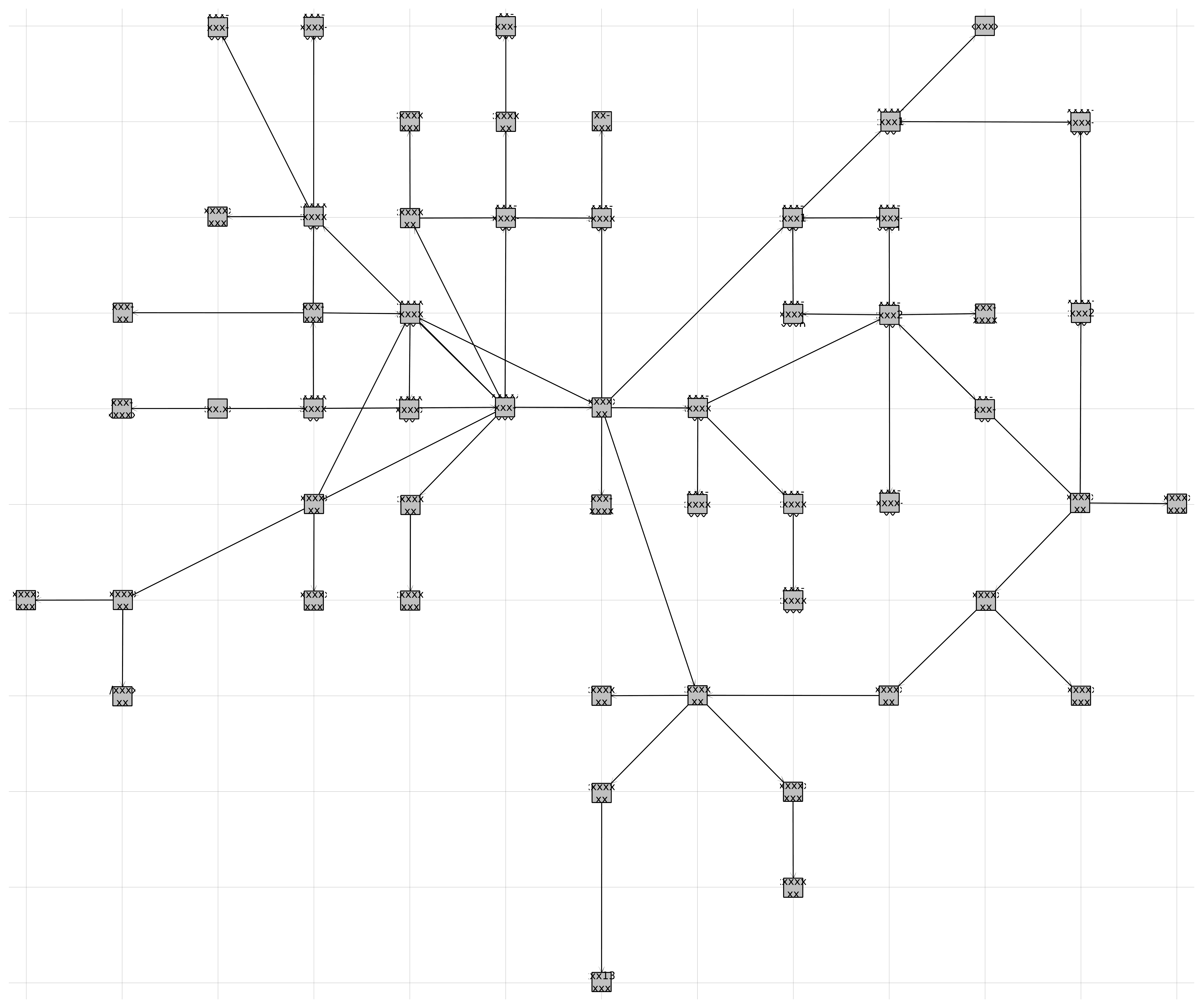}
}
\subfigure[ACA]{
\includegraphics[width=0.3\textwidth]{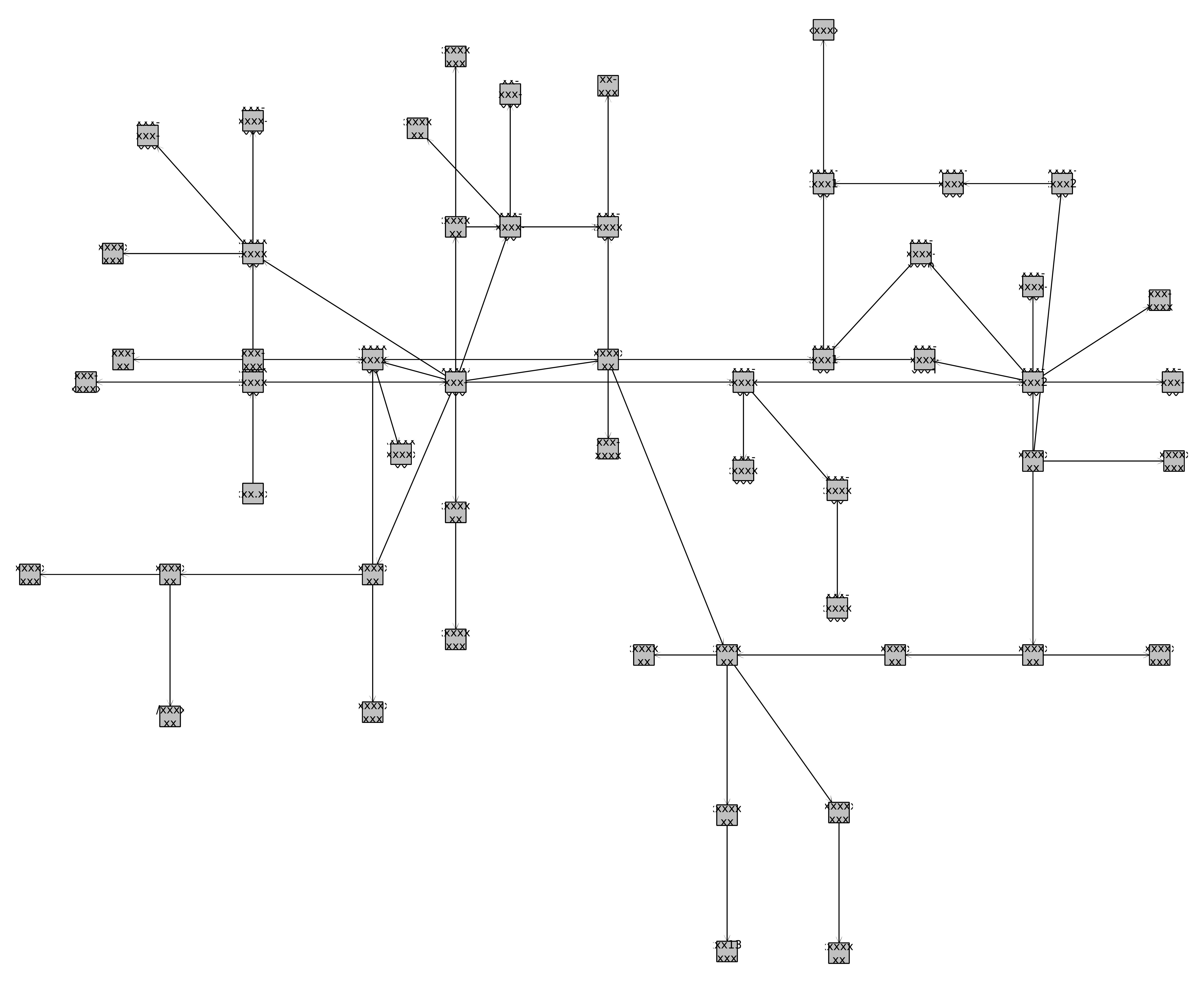}
}
\subfigure[ACA + GS]{
\includegraphics[width=0.3\textwidth]{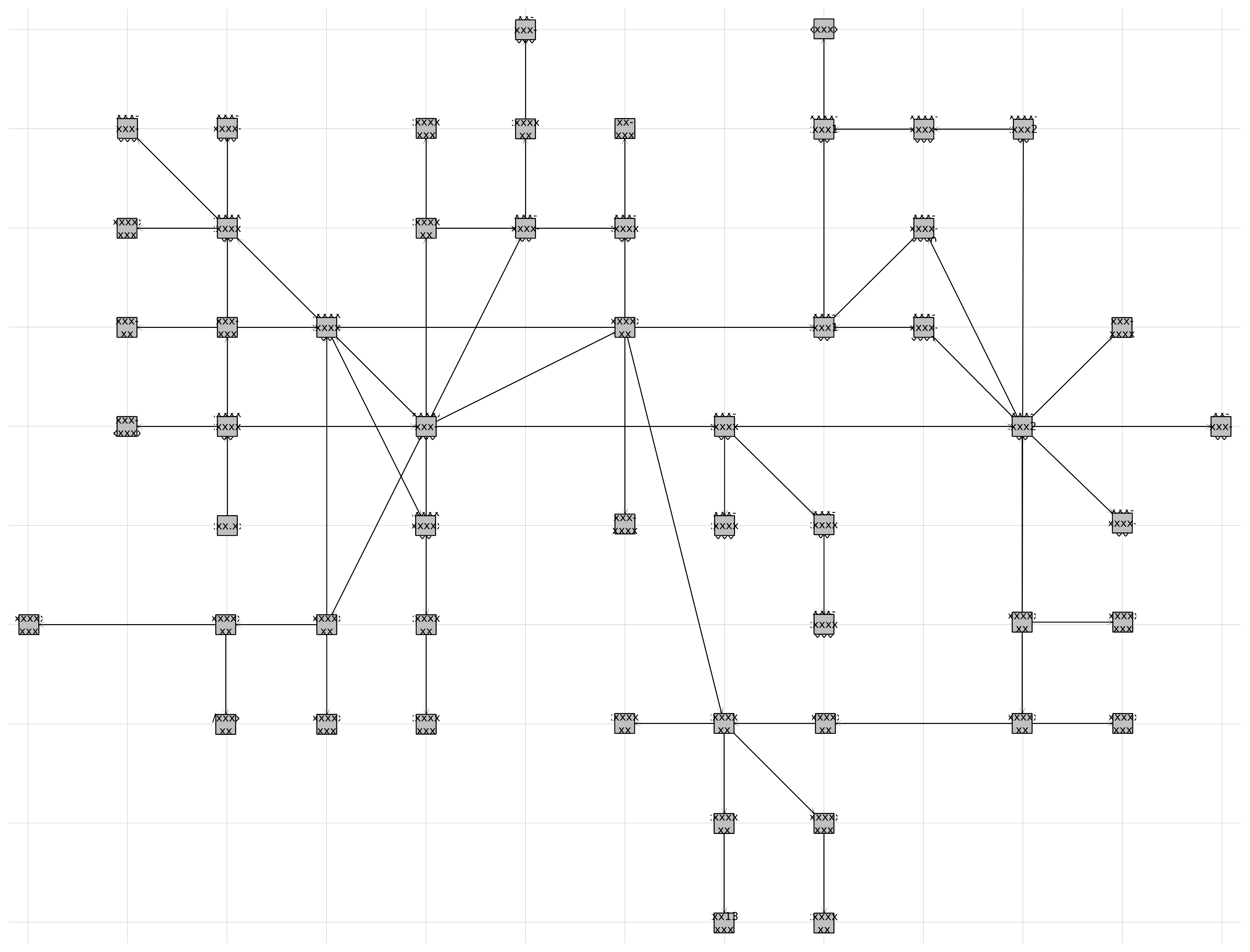}
}
\caption{Different combinations of our automatic layout techniques for the graph ``ug\_22'' from the AT\&T Graphs corpus, as generated during our evaluation.}
\label{fig:eval-22}
\end{figure}

We give here more detailed results from our experimental application of our six methods to the AT\&T Graphs corpus, as described in Section 6 of the paper.  In the figures here (Figures~\ref{fig:eval-time}--\ref{fig:eval-stress}), we refer to: unconstrained (except for non-overlap constraints) Force Directed layout as FD, Grid-Snap as GS, Node-Snap as NS, and Adaptive Constrained Alignment as ACA.  The experiment was run on a MacBook Pro with a 2.3GHz Intel Core I7 CPU.  Output for three of the graphs, for each of the six methods, can be seen in Figures~\ref{fig:eval-213}--\ref{fig:eval-22}.

\end{document}